\documentclass[a4paper,twocolumn,11pt,accepted=2024-06-18]{quantumarticle}
\pdfoutput=1
\usepackage[utf8]{inputenc}
\usepackage[english]{babel}
\usepackage[T1]{fontenc}
\usepackage{amsmath}
\usepackage{hyperref}
\usepackage{tikz}
\usetikzlibrary{quantikz2}
\usepackage{lipsum}

\usepackage[numbers]{natbib}

\usepackage{graphicx}
\usepackage{dcolumn}
\usepackage{amssymb}

\begin{document}

\title{Estimation of high-dimensional unitary transformations saturating the Quantum Cram\'er-Rao bound}

\author{J. Escand\'on-Monardes}
\email{jescandon@udec.cl}
\author{D. Uzc\'ategui}
\author{M. Rivera-Tapia}
\author{S. P. Walborn}
\author{A. Delgado}
\affiliation{Departamento de F\'{\i}sica, Universidad de Concepci\'on, 160-C Concepci\'on, Chile}
\affiliation{Millennium Institute for Research in Optics, Universidad de Concepci\'on, 160-C Concepci\'on, Chile}

\maketitle

\begin{abstract}
We propose an estimation procedure for $d$-dimensional unitary transformations. For $d>2$, the unitary transformations close to the identity are estimated saturating the quantum Cramér-Rao bound. For $d=2$, the estimation of all unitary transformations is also optimal with some prior information.  We show through numerical simulations that, even in the absence of prior information, two-dimensional unitary transformations can be estimated with greater precision than by means of standard quantum process tomography.
\end{abstract}

\section{Introduction}

The continuous advancement in the ability to control quantum systems and its application to the development of quantum technologies has driven the search for high-precision measurements and estimation methods. Quantum metrology aims to develop methods and tools that achieve the ultimate precision in parameter estimation. The enhancements provided by quantum metrology depend on the state of the probe, the quantum measurement, and the landscape of the parameters to be estimated. These are usually related through a multivariate variational problem that generally lacks analytical solutions. Despite this, quantum metrological improvements have already been demonstrated on various experimental platforms in both the single parameter and multiparameter cases \cite{Nagata,Higgins,Xiang,Slussarenko,Daryanoosh,Hou,walborn18,Vidrighin,Roccia_2018,Hou_etal_2018,Conlon_etal_2023}.

Due to its intrinsic difficulty \cite{Liu_2020}, the multiparameter case has remained less explored. In particular, the optimal measurements for different parameters are often incompatible \cite{Vidrighin} and the optimal probe states for different parameters can typically be different. Furthermore, in the multiparameter estimation, the quantum Cram\'er-Rao bound \cite{Helstrom,Holevo}, which sets a fundamental limit for the covariance matrix, is generally not achievable even asymptotically \cite{Humphreys,Ragy2016}. However, there are other Cram\'er-Rao type bounds, which are defined considering a weighted trace of the covariance matrix and are attainable in some scenarios \cite{GillMassar,Conlon_etal_2021,Hayashi2023,Demkowicz2020,Albarelli2020}.

An instance of multiparamenter estimation is the estimation of $d$-dimensional unitary transformations. Several methods to accomplish this task have been studied \cite{balwin_2014, Zhou_2015}, particularly standard quantum process tomography (SQPT) \cite{chuang_1997}, which has been successfully implemented for reconstructing quantum gates on ion traps \cite{Riebe_2006}, superconducting circuits \cite{bialczak_2010}, among many others \cite{childs_2001,Howard_2006}.

Here, we propose a novel method for estimating $d$-dimensional unitary operations. Our approach requires a single target qudit, two control qudits, controlled gates, and Fourier transformations acting on the control qudits. The unknown unitary transformation acts on the target qudit. These resources allow mapping the unitary transformation to a state of both control qudits, which, after performing measurements, lead to an estimate of the coefficients that define the unitary transformation in the Weyl-Heisenberg basis. This is achieved without the need to measure the target qudit. Although our procedure does not provide a quantum metrological advantage, we show that it saturates the quantum Cram\'er-Rao bound for any initial target qudit state and simply uses measurements in the computational basis. This result holds for any finite dimension $d>2$ and for unitary transformations that are close to the identity. In the case $d=2$, all unitary transformations  can be optimally estimated provided that the octant which the Bloch vector points to is known in advance. In this case, our estimation procedure agrees with previous results \cite{Ballester_2004,yuan_2016,hou_etal_2021}. In both cases the estimation scheme is independent of the parameters of the unitary being estimated.

We also estimate $2$-dimensional unitary transformations without prior information, at the cost of losing estimation accuracy, and compare with SQPT. We simulate both procedures on Qiskit, IBM's software development platform for quantum processors \cite{Qiskit}, and study the average gate fidelity \cite{NIELSEN2002249} as a function of the ensemble size, or number of shots, for a set of randomly generated unitary transformations. In the ideal case, that is, in the absence of error sources, our estimation procedure provides a better mean average gate fidelity than SQPT. Moreover, our estimation procedure shows a narrow standard deviation which means that all unitary transformations are estimated with similar average gate fidelity. In presence of noise affecting state generation, quantum gates, and measurements, our estimation procedure and SQPT exhibit similar mean gate infidelity, although the former with a larger standard deviation. Median gate fidelity of our procedure is larger than that of SQPT, which lays close to the inferior border of the interquartile range of our estimation procedure. Thereby, our estimation procedure provides a better estimation than SQPT in most cases.

\section{Preliminary Material}
\subsection{Unitary transformations}

An arbitrary $d$-dimensional unitary transformation $U$ can be expanded in the Weyl-Heisenberg basis as
\begin{equation}
U=\sum_{m,n=0}^{d-1}u_{m,n}X^mZ^n,
\label{Weyl-Heisenberg expansion}
\end{equation}
where $X$ and $Z$ are the shift and phase operators, respectively. These act onto the canonical basis $\{|k\rangle\}$ with $k=0,\dots,d-1$ as
$X|k\rangle=|k\oplus 1\rangle$ and $Z|k\rangle=\omega^k|k\rangle$ with $\omega=\exp(2i\pi/d)$. The set $\{u_{m,n}=r_{m,n}e^{i\phi_{m,n}}\}$ of $d^2$ complex coefficients satisfies the unitarity constraint and characterizes $U$.
With this, a general unitary can be written
\begin{equation}
U=r_{0,0} I + \sum \limits_{ \substack{m,n=0 \\ (m,n)\neq (0,0)}}^{d-1}r_{m,n}e^{i\phi_{m,n}}X^mZ^n.
\label{Weyl-Heisenberg expansion_rphi}
\end{equation}
where we set $\phi_{0,0}=0$ without loss of generality.

\subsection{Quantum estimation theory}
\label{Quantum estimation theory}

The classical Cramér-Rao bound states that the covariance matrix $cov(\hat{\bf t})$ of an unbiased estimator $\hat{\bf t}$ of a parameter vector ${\bf t}$ is bounded below by the inverse of Fisher information matrix $\mathcal{I}(\bf t)$, that is, for $n$ repetitions of the experiment,
\begin{equation}
cov(\hat{\bf t})\ge\frac{1}{n}\mathcal{I}^{-1}(\bf t),
\label{Classical Cramer-Rao}
\end{equation}
which leads to limits for the accuracy of the estimate under various figures of merit (for recent reviews on the topic, see Refs. \cite{Liu_2020} and \cite{barbieri_2022}). The entries of the Fisher information matrix are defined as
\begin{equation}
    \mathcal{I}_{ab}=\sum_y \frac{1}{P(y|{\bf t})}\left[\frac{\partial P(y|{\bf t})}{\partial t_a}\right]\left[\frac{\partial P(y|{\bf t})}{\partial t_b}\right],
\end{equation}
where $P(y|{\bf t})$ is the probability of observing the value $y$ in an experiment for a given parameter vector ${\bf t}$. In the case of quantum mechanics the probability distribution $P(y|{\bf t})$ depends on the measurement performed, which leads to different Fisher information matrices. Moreover, it is possible to derive (see Appendix H in Ref. \cite{Liu_2020}) a new bound known as the quantum Cramér-Rao bound given by
\begin{equation}
cov(\hat{\bf t})\ge\frac{1}{n}\mathcal{I}^{-1}({\bf t})\ge\frac{1}{n}\mathcal{F}^{-1}({\bf t}),
\label{Quantum Cramer-Rao}
\end{equation}
where $\mathcal{F}$ is the quantum Fisher information matrix. Eq.\thinspace(\ref{Quantum Cramer-Rao}) sets a bound to the achievable precision in the estimation of a set of parameters in the context of Quantum Mechanics. In the case of estimating a unitary transformation that acts onto a probe state $|\phi\rangle$, the quantum Fisher information matrix can be calculated as  \cite{Liu_2020}
\begin{equation}
\mathcal{F}_{a,b}=2\langle\phi|\{H_a,H_b\}|\phi\rangle-4\langle\phi|H_a|\phi\rangle\langle\phi|H_b|\phi\rangle,
\label{Quantum Fisher matrix definition}
\end{equation}
where we have $H_a=i(\partial_aU^\dagger)U$.

\section{Results}
\label{Results}

\subsection{Estimation procedure}
\label{Optimal estimation procedure}

\begin{figure*}[t]
  \centering
  \begin{quantikz}[column sep=0.4cm]
      \lstick[wires=1]{$\ket{\psi}_0$} \slice{$|\Phi^0\rangle_{012}$} & \qw & \gate[wires=1][0.8cm][0.8cm]{X_{02}^{(0)}} & \gate[wires=1][0.8cm][0.8cm]{Z_{01}^{\dagger(1)}} & \gate[wires=1][0.8cm][0.8cm]{U} & \gate[wires=1][0.8cm][0.8cm]{Z_{01}^{(0)}} & \gate[wires=1][0.8cm][0.8cm]{X_{02}^{\dagger(1)}} & \qw \slice{$|\Phi^7\rangle_{012}$} & \gate[wires=1][0.8cm][0.8cm]{X_{01}^{\dagger(-1)}} & \gate[wires=1][0.8cm][0.8cm]{Z_{02}^{\dagger(0)}} & \qw \slice{$|\Phi^{10}\rangle_{012}$} & \qw \rstick[wires=1] {$\ket{\psi}_0$} \\
      \lstick[wires=1]{$\ket{0}_1$} & \gate[wires=1][0.8cm][0.8cm]{F} & \qw & \ctrl{-1} & \qw & \ctrl{-1} & \qw & \gate[wires=1][0.8cm][0.8cm]{F^{-1}} & \ctrl{-1} & \qw & \qw & \qw \rstick[wires=2]{$\ket{\varphi^0}_{12}$}\\
      \lstick[wires=1]{$\ket{0}_2$} & \gate[wires=1][0.8cm][0.8cm]{F} & \ctrl{-2} & \qw & \qw & \qw & \ctrl{-2} & \gate[wires=1][0.8cm][0.8cm]{F^{-1}} & \qw & \ctrl{-2} & \gate[wires=1][0.8cm][0.8cm]{X} & \qw 
  \end{quantikz}
  \caption{Quantum circuit implementation of our estimation procedure. $F$ are d-dimensional Fourier transforms acting on control qudits 1 and 2, $X^{(i)}_{tc}$ and $Z^{(i)}_{tc}$ are controlled gates defined in Eq.~\eqref{ControlledGates}, and $U$ is the unitary transformation to be estimated.  
  State tomography of the control system leads to complete estimation of $U$.}
  \label{Circuit0}
\end{figure*}
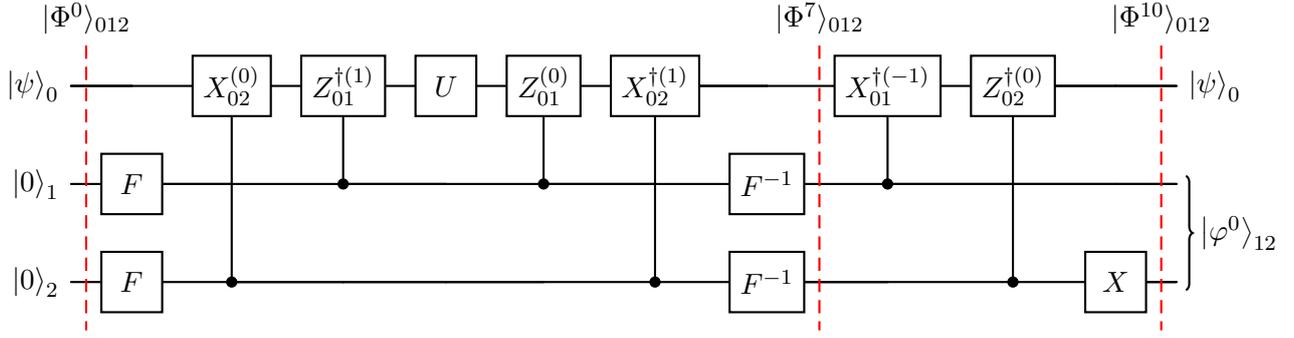

 To estimate a $d$-dimensional unitary operation  $U$, we propose a procedure which uses the quantum circuit shown in Fig. \ref{Circuit0}. This circuit is applied to the following initial quantum state of three qudits
\begin{equation}
|\Phi^0\rangle_{012} = |\psi\rangle_0\otimes |00\rangle_{12},
\label{State0}
\end{equation}
where the target qudit is in the arbitrary state $|\psi\rangle_0$, and the qudits labeled $1$ and $2$ are the control states. The superscript $i$ in $|\Phi^i\rangle_{012}$ indicates the time step in the circuit. Each control qudit is subject to the action of a Fourier transform $F|k\rangle=(1/\sqrt{d})\sum\omega^{mk}|m\rangle$, followed by the sequence of controlled shift and phase operators $X_{02}^{(0)}$ and $Z_{01}^{\dagger(1)}$  defined by
\begin{align}
\label{ControlledGates}
X_{tc}^{(i)}&=\sum_{m=0}^{d-1}X_t^m \otimes|m\ominus i\rangle_c\langle m\ominus i|,\nonumber\\
Z_{tc}^{(i)}&=\sum_{m=0}^{d-1}Z_t^m\otimes |m\ominus i\rangle_c\langle m\ominus i|.
\end{align}
The action of the previous transformations leads to the probe state
\begin{equation}
\ket{\Phi^3}_{012}=\frac{1}{d}\sum_{j_1,j_2=0}^{d-1}Z_0^{-j_1-1}X_0^{j_2}\ket{\psi}_0\otimes\ket{j_1}_1\otimes\ket{j_2}_2.
\end{equation}
 The unitary transformation $U$ to be estimated acts on the target qudit followed by the sequence $Z_{01}^{(0)}$ and then $X_{02}^{\dagger(1)}$ and inverse Fourier transforms acting on each control qudit. Thereby, the initial state is transformed into the state
\begin{equation}
|\Phi^7\rangle_{012}=\sum_{m,n=0}^{d-1}u_{m,n+1}(X^{m-1}Z^n|\psi\rangle_0)\otimes|m\rangle_1 \otimes|n\rangle_2,
\end{equation}
then $X_{01}^{\dagger(-1)}$ and $Z_{02}^{\dagger(0)}$ disentangles the target qudit from the control qudits, followed by $X_2$ which correlates the indexes of the coefficients with the computational basis of both control qudits. The final state (see Appendix \ref{Final_state} for details) is
\begin{equation}
|\Phi^{10}\rangle_{012}=|\psi\rangle_0\otimes|\varphi^0\rangle_{12},
\end{equation}
where
\begin{equation}
\ket{\varphi^0}_{12} = \sum_{m,n=0}^{d-1}u_{m,n}|m\rangle_1\otimes|n\rangle_2.
    \label{varphi0_main}
\end{equation}
The coefficients $u_{m,n}$ entering in Eq.~\eqref{Weyl-Heisenberg expansion} are now in a one-to-one relation with the states in the canonical basis of the control qudits. Thus, the estimation of state $\ket{\varphi^0}_{12}$ by any quantum tomographic scheme for pure states leads to the estimation of the unknown unitary transformation $U$. Moreover, simpler measurement schemes can be used, provided that a pure state of two qudits is defined by a set of $2d^2-2$ independent real parameters while a unitary transformation acting on a single qudit is characterized only by $d^2-1$ real parameters. Some specific cases are studied below.

\subsection{Estimation of 2-dimensional unitary transformations}
\label{2-dimensional case}

Let us now consider the case of $2$-dimensional unitary transformations. These can be written as \cite{nielsen_chuang_2010}
\begin{equation}
U=\exp(-i\alpha \hat{n}\cdot \hat{\sigma}),
\label{u_main}
\end{equation}
where $\hat{\sigma}=(X,Y,Z)^T$ is the Pauli vector, $\alpha\in[0,\pi/2]$, and $\hat{n}\in\mathbb{R}^3$ is a real unitary vector. After carrying out the exponentiation, we obtain the representation 
\begin{equation}
\label{U_Pauli_decomposed}
U=u_{0,0}I+u_{1,0}X+u_{1,1}XZ+u_{0,1}Z,
\end{equation}
where we replaced $Y = iXZ$, and the coefficients are given by
\begin{align}
        u_{0,0} &= \cos(\alpha), \nonumber\\
        u_{1,0} &= -i\sin(\alpha)\sin(\theta)\cos(\phi),\nonumber\\
        u_{1,1} &= \sin(\alpha)\sin(\theta)\sin(\phi),\nonumber\\
        u_{0,1} &= -i\sin(\alpha)\cos(\theta),
        \label{Relations_u}
\end{align}
with $\theta\in[0,\pi]$ and $\phi\in[0,2\pi[$ being the spherical coordinates for $\hat{n}$, and $I$ the $2\times2$ identity matrix. Notice that $u_{0,0}$ is always non-negative, whereas the signs of $u_{1,0}, u_{1,1}, u_{0,1}$ depend on the octant in which the vector $\hat{n}$ points. Estimating an unknown two-dimensional unitary transformation $U$ is thus equivalent to estimating the values of the angles $(\alpha, \theta, \phi)$. Projective measurements of control qudits lead to probabilities
\begin{align}
    P_{0,0} &=\cos^2(\alpha),\\
    P_{1,0} &=\sin^2(\alpha)\sin^2(\theta)\cos^2(\phi),\\
    P_{1,1} &=\sin^2(\alpha)\sin^2(\theta)\sin^2(\phi),\\
    P_{0,1} &=\sin^2(\alpha)\cos^2(\theta).
    \label{probabilities_circ_0}
\end{align}
It follows that
\begin{align}
    \cos^2(\alpha)&=P_{0,0},\nonumber\\ 
    \cos^2(\theta)&=\frac{P_{0,1}}{1-P_{0,0}},\nonumber\\
    \cos^2(\phi)&=\frac{P_{1,0}}{P_{1,1}+P_{1,0}}.
    \label{FirstEstimate}
\end{align}

These relations allow for estimating the value of $\alpha$, which is always in the interval $[0,\pi/2]$. However, parameters $\theta$ and $\phi$ remain ambiguous, since $u_{0,1}, u_{1,0}, u_{1,1}$ are determined up to a sign. This ambiguity is removed when the octant pointed to by $\hat{n}$ is known beforehand, in which case our estimation procedure characterizes the unknown unitary transformation. Furthermore, it can be shown by direct algebra that our estimation procedure fulfills the equality $\mathcal{I}=\mathcal{F}$ where
\begin{equation}
\label{QFIMQUBIT}
\mathcal{F}=4\begin{pmatrix}
    1 & 0 & 0\\
    0 & \sin^2(\alpha) & 0\\
    0 & 0 & \sin^2(\alpha)\sin^2(\theta)
\end{pmatrix},
\end{equation}
and therefore our proposal saturates the quantum Cramér-Rao bound. Moreover, $\mathcal{F}$ is diagonal, hence our circuit is optimal in the sense that the three parameters defining $U$ can be estimated simultaneously with the highest possible precision. The quantum Fisher information matrix in Eq.~(\ref{QFIMQUBIT}) was also obtained in other works \cite{Ballester_2004,yuan_2016,hou_etal_2021}.

The lack of a priori information does not prevent the use of our estimation procedure. As we show through numerical simulations in section \ref{sec:num}, our procedure can be complemented with additional measurements and at the same time achieves better estimation accuracy than that obtained by standard process tomography.

\subsection{Estimation of higher-dimensional unitary transformations}
\label{d-dimensional case}

In order to estimate a higher-dimensional unitary transformation we need a clear dependence of $U$ in terms of $d^2-1$ independent real parameters. For example, $U$ can be parametrized in a similar fashion to Eq.\thinspace\eqref{u_main} as $U=\exp\left( i \sum_{j=1}^{d^2-1} \lambda_j T_{j} \right)$, where $T_{j}$ are the generalized Gell-Mann matrices and $\lambda_j$ are the independent real parameters. For $d=2$ we expanded this expression in the Weyl-Heisenberg basis as in Eq.\thinspace{\eqref{U_Pauli_decomposed}}, where the complex coefficients $u_{m,n}$ explicitly depend on the corresponding parameters $\lambda_j$ (see Eq.\thinspace\eqref{Relations_u}). However, it is not possible to obtain equivalent relations for $d>2$ analytically, leading to difficulties in both the estimation of $\lambda_j$ and the calculation of the classical and quantum Fisher information matrices. Furthermore, finding a convenient parametrization that facilitates these calculations is an open problem. In this work, we consider the expansion of $U$ from Eq.\thinspace\eqref{Weyl-Heisenberg expansion_rphi} where the elements in $\{ r_{m,n}, \phi_{m,n} \}$ are not independent from each other.  We restrict the analysis to unitary transformations that are close to the identity, since this approximation uncouples the parameters and allows us to obtain analytically both the classical and quantum Fisher information matrices.

To simplify the notation, we denote the coefficients $u_{p_x,p_z} \equiv u_p=r_pe^{i\phi_p}$, where we have defined a single index $p=(p_x,p_z)$. These coefficients are constrained by the conditions
\begin{equation}
\label{U-normalization}
    \sum_{m\in\mathbb{Z}_d^2}r_m^2=1
\end{equation}
and
\begin{equation}
\label{Unitarity-condition}
    \sum_{m\in\mathbb{Z}_d^2}r_mr_{p\oplus m}e^{i(\phi_{p\oplus m}-\phi_m)} \omega^{-m_xp_z}=0,~\forall~p\neq(0,0),
\end{equation}
which enforce unitarity (see Appendix \ref{App:Unitarity}). For unitary transformations close to the identity we have that $r_m/r_0\ll1$ for $m\ne(0,0)$. With this approximation, the only terms contributing in Eq.~\eqref{Unitarity-condition} are those where $m=(0,0)$ and $m=\ominus p\equiv(d-p_x,d-p_z)$. Thus, Eq.~\eqref{Unitarity-condition} becomes
\begin{equation}
    r_{p}e^{i\phi_p} \approx r_{\ominus p}e^{-i\phi_{\ominus p}+i\frac{2\pi}{d}p_xp_z+i\pi}  ,
\end{equation}
where we set $\phi_{0,0}=0$ without loss of generality. From the previous equation we obtain
\begin{equation}
    \label{paired-amplitud}
    r_p \approx r_{\ominus p}
\end{equation}
and
\begin{equation}
    \label{paired-phase}
    \phi_p \approx -\phi_{\ominus p}+\frac{2\pi}{d}p_xp_z+(2n+1)\pi\;,\text{with }n\in\mathbb{Z}.
\end{equation}
These conditions show that the amplitudes and phases of the coefficients $u_p$ are related in pairs. In the case $p=\ominus p$, Eq.~\eqref{paired-phase} ties the phase to a discrete set, that is,
\begin{equation}
    \phi_p \approx \frac{\pi}{d}p_xp_z+\frac{2n+1}{2}\pi\;,\text{with }n\in\mathbb{Z}.
    \label{fixed-phase}
\end{equation}
Notice that the last restriction on the phases only occurs when $d$ is even, and only for $p=(d/2,0)$, $p=(0,d/2)$ and $p=(d/2,d/2)$. Therefore, in the case $d=2$, the three coefficients have a fixed phase up to a difference of $\pi$.

The constraints in Eqs. \eqref{paired-amplitud} and \eqref{paired-phase} allow us to recognize the $d^2-1$ parameters that characterizes $U$ in the close-to-the-identity approximation. These are, for $d=2$, three unpaired amplitudes $r_p$. For $d>2$ odd, all the coefficients are paired, implying that the relevant parameters are $(d^2-1)/2$ amplitudes $r_p$ and the same number of phases $\phi_p$. For $d>2$ even, we have the three unpaired amplitudes $r_{(d/2,0)}$, $r_{(0,d/2)}$ and $r_{(d/2,d/2)}$, and $(d^2-4)/2$ other amplitudes and equal number of phases. 

To handle these three cases at once we introduce the following partition of the set $\mathbb{Z}_d^2$ of indexes:
\begin{equation}
    \mathbb{Z}_d^2=S_0\cup S_u\cup S_+\cup S_{-}\;,
    \label{partition_main}
\end{equation}
where $S_0=\{(0,0)\}$, $S_u=\{(d/2,0),(0,d/2),(d/2,d/2)\}$, and $S_+$ and $S_-$ are any partition such that $p\in S_+$ if and only if $\ominus p\in S_-$. In this way, we can easily identify the set of parameters defining $U$:
\begin{equation}
    \texttt{Par}_U=\{r_f\}_{f\in S_u}\cup\{r_a,\phi_a\}_{a\in S_+}\;.
\end{equation}
Notice that every $r_a$ and $\phi_a$ with $a\in S_+$ is respectively paired with $r_{\ominus a}$ and $\phi_{\ominus a}$, with $\ominus a\in S_-$, via Eqs. \eqref{paired-amplitud} and \eqref{paired-phase}. Here and in what follows we use indexes $f,g\in S_u$ to label unpaired amplitudes and $a,b\in S_+$ to label paired amplitudes and phases. 

Let us introduce the notation $\ket{f}=\ket{f_x}_1\otimes\ket{f_z}_2$ for index $f=(f_x,f_z)$, and similarly for $a=(a_x,a_z)$. Then, using partition \eqref{partition_main}, the state $\ket{\varphi^0}_{12}$ in Eq.~\eqref{varphi0_main} becomes
\begin{align}
    \ket{\varphi^0}_{12} =&\; r_0\ket{0} +  \sum_{f\in S_u}r_fe^{i\phi_f}\ket{f}\nonumber\\
    &+\sum_{a\in S_+}r_a\left(e^{i\phi_a}\ket{a}+e^{i\phi_{\ominus a}}\ket{\ominus a}\right),
\end{align}
where  $r_0 = (1 - \sum_{n\neq(0,0)}r_n^2)^{1/2}$.

Now, consider the following two-qudit operation
\begin{equation}
    \tilde{H}\ket{n}=\begin{cases}
        \ket{n} & \text{, for $n=0$ and $n\in S_u$.}\\
        \frac{1}{\sqrt{2}}\left(\ket{n}+\ket{\ominus n}\right) & \text{, for $n\in S_+$.}\\
        \frac{1}{\sqrt{2}}\left(\ket{\ominus n}-\ket{n}\right) & \text{, for $n\in S_-$.}
    \end{cases}
\end{equation}
This operation can be understood as a set of Hadamard gates each acting in a subspace labeled with paired indexes, while acting as an identity on the other subspaces. Applying $\tilde{H}$ on $\ket{\varphi^0}_{12}$ we obtain the state
\begin{align}
    \ket{\varphi^1}_{12}=& \; r_0\ket{0} +  \sum_{f\in S_u}r_fe^{i\phi_f}\ket{f}\nonumber\\ 
        &+ \sum_{a\in S_+}\frac{r_a}{\sqrt{2}}\left(e^{i\phi_a}+e^{i\phi_{\ominus a}}\right)\ket{a}\nonumber\\ 
        &+ \sum_{a\in S_+}\frac{r_a}{\sqrt{2}}\left(e^{i\phi_a}-e^{i\phi_{\ominus a}}\right)\ket{\ominus a}.
\end{align}
Projective measurements on the computational basis for both qudits leads to the probabilities
\begin{align}
    P_0 &= r_0^2,\nonumber\\
    P_f &= r_f^2,\nonumber\\
    P_a &= r_a^2(1+\cos(\Delta_a)),\nonumber\\
    P_{\ominus a} &= r_a^2(1-\cos(\Delta_a)),
    \label{probs_main}
\end{align}
where $\Delta_a= \phi_a-\phi_{\ominus a}$ is given by the expression
\begin{equation}
    \Delta_a=2\phi_a-\frac{2\pi}{d}a_xa_z-(2n+1)\pi\;, \text{with $n\in\mathbb{Z}$.}
    \label{Delta_a_main}
\end{equation}
These probabilities lead to the estimates for the amplitudes
\begin{align}
    r_f &= \sqrt{P_f}\nonumber\\
    r_a &=\sqrt{\frac{P_a+P_{\ominus a}}{2}},
\end{align}
and for the phases
\begin{align}
    \phi_a =& \pm\frac{1}{2}\arccos\left(\frac{P_a-P_{\ominus a}}{P_a+P_{\ominus a}}\right)+\frac{\pi}{d}a_xa_z\nonumber\\
    &+(n+\frac{1}{2})\pi \;, \text{with $n\in\mathbb{Z}$.}
    \label{estimated_phases}
\end{align}
Thus, our proposal estimates the amplitudes and phases that characterize any $d$-dimensional close-to-the-identity unitary gate. In any case, the phases are estimated up to a set of four candidates, as implied by Eq. \eqref{estimated_phases} and in agreement with the 2-dimensional case. The discrimination of the candidates requires prior information or additional experiments.

For dimension $d$ even, the quantum Fisher information matrix characterizing our process is given (see Appendix \ref{Calculation of Fisher Information matrix} for details) by the block matrix
\begin{equation}
    \mathcal{F}_{even} = \begin{pmatrix}
        \mathbf{A} & \mathbf{B} & \mathbf{0}\\
        \mathbf{B}^T & \mathbf{C} & \mathbf{0}\\
        \mathbf{0} & \mathbf{0} & \mathbf{D}
    \end{pmatrix},
    \label{F_even_main}
\end{equation}
where the ordering in the block matrix $\mathcal{F}_{even}$ is given by $(\{r_f\},\{r_a\},\{\phi_a\})$, with $f\in S_u$ and $a \in S_{+}$, and the explicit form of each block is
\begin{align}
    \mathbf{A}_{f,g} &= 4\frac{r_fr_g}{r_0^2} + 4\delta_{f,g}\\
    \mathbf{B}_{f,a} &= 8\frac{r_fr_a}{r_0^2}\\
    \mathbf{C}_{a,b} &= 16\frac{r_ar_b}{r_0^2} + 8\delta_{a,b}\\
    \mathbf{D}_{a,b} &= 8r_a^2 \delta_{a,b}\;,
\end{align}
being $\delta_{x,y}$ the Kronecker delta. In particular, for $d=2$, the unitary transformation is characterized by three unpaired amplitudes, i.e., $S_+$ and $S_-$ are empty. Hence, the quantum Fisher information matrix reduces to the upper left block as
\begin{equation}
    \mathcal{F}_{2} = \mathbf{A}\;.
    \label{F_2}
\end{equation}
In the case of dimension $d$ odd all amplitudes and phases are paired, hence $S_u$ is empty. Thus, the quantum Fisher information matrix is given by
\begin{equation}
    \mathcal{F}_{odd} = \begin{pmatrix}
        \mathbf{C} & \mathbf{0}\\
        \mathbf{0} & \mathbf{D}
    \end{pmatrix}.
    \label{F_odd}
\end{equation}

In this way, we have obtained the quantum Fisher information matrix $\mathcal{F}_{d}$ for estimating close-to-the-identity unitary transformations in every dimension. Furthermore, we show in Appendix \ref{d_dimensional_Classic_Fisher_Appendix} that the classical Fisher information matrix $\mathcal{I}_{d}$ is equal to $\mathcal{F}_{d}$. Thus, our estimation procedure saturates the quantum Cram\'er-Rao inequality for close-to-the-identity unitary transformations. Let us note that within the approximation $r_m/r_0\ll1$ the non-diagonal terms in $\mathcal{F}_{even}$ are $O((r_m/r_0)^2)$, hence they can be neglected and consequently all the Fisher information matrices are nearly diagonal. In this way, in the case of odd dimension, the amplitudes can be estimated independently of each other and with equal precision. For even dimension, the amplitudes are also estimated independently; however, unpaired amplitudes are estimated with half the precision of paired amplitudes. Lastly, the precision in the estimation of the phases is severely restricted since it is proportional to the inverse of the square of the corresponding amplitude.

We must emphasize that our result is only valid within the close-to-the-identity approximation, where the parameters in the Weyl-Heisenberg expansion of $d$-dimensional unitary matrices, namely the phases $\{ \phi_{p} \}$ and the amplitudes $\{ r_{p} \}$, are approximately independent between each other. In the case far from the identity, where the parameters are strongly correlated, the above does not hold: even a slight variation in one of the amplitudes, which produces slight variations in the other ones because of the normalization, leads to violation of the unitarity conditions. Since the relation between the parameters for the general case is not clear, it is not possible to calculate, even numerically, the partial derivatives required to obtain the classical and quantum Fisher information matrices. In Appendix \ref{AppX} we use a different parametrization to compare these matrices, showing that the distance between them increases as $U$ is further from the identity.

\subsection{Numerical simulations}
\label{sec:num}

In this section we study the performance of our estimation procedure for the case of qubit gates without prior information. This is achieved by measuring the quantum state in Eq. \eqref{varphi0_main} in three different bases. The resulting statistics completely characterizes the unknown unitary transformation, as shown in Appendix \ref{no_prior_information}.

We simulate our estimation procedure with Qiskit \cite{Qiskit}, IBM's software development platform for quantum processors, and compare its performance against the built-in function for SQPT. Since the output of SQPT is a quantum channel that is not necessarily unitary, contrary to our procedure which guarantees unitarity,  we use as figure of merit the average gate fidelity, which compares a general quantum channel with a unitary quantum transformation \cite{NIELSEN2002249}. We generate a set of $200$ single-qubit unitary matrices, which are randomly drawn from a uniform Haar distribution. Each unitary transformation is reconstructed 1000 times using our estimation procedure and additionally SQPT. In both strategies we calculate the average gate fidelity respect to the ideal unitary gate. This is repeated using increasing number of shots (or ensemble sizes) to simulate the measurement results. Finally, we also perform simulations considering various error sources (see Appendix \ref{app:noise}) and readout error mitigation. The code implementing this method is available in a Github repository \cite{github}.

\begin{figure*}[t]
\centering
\begin{tabular}{cc}
\includegraphics[width=70mm]{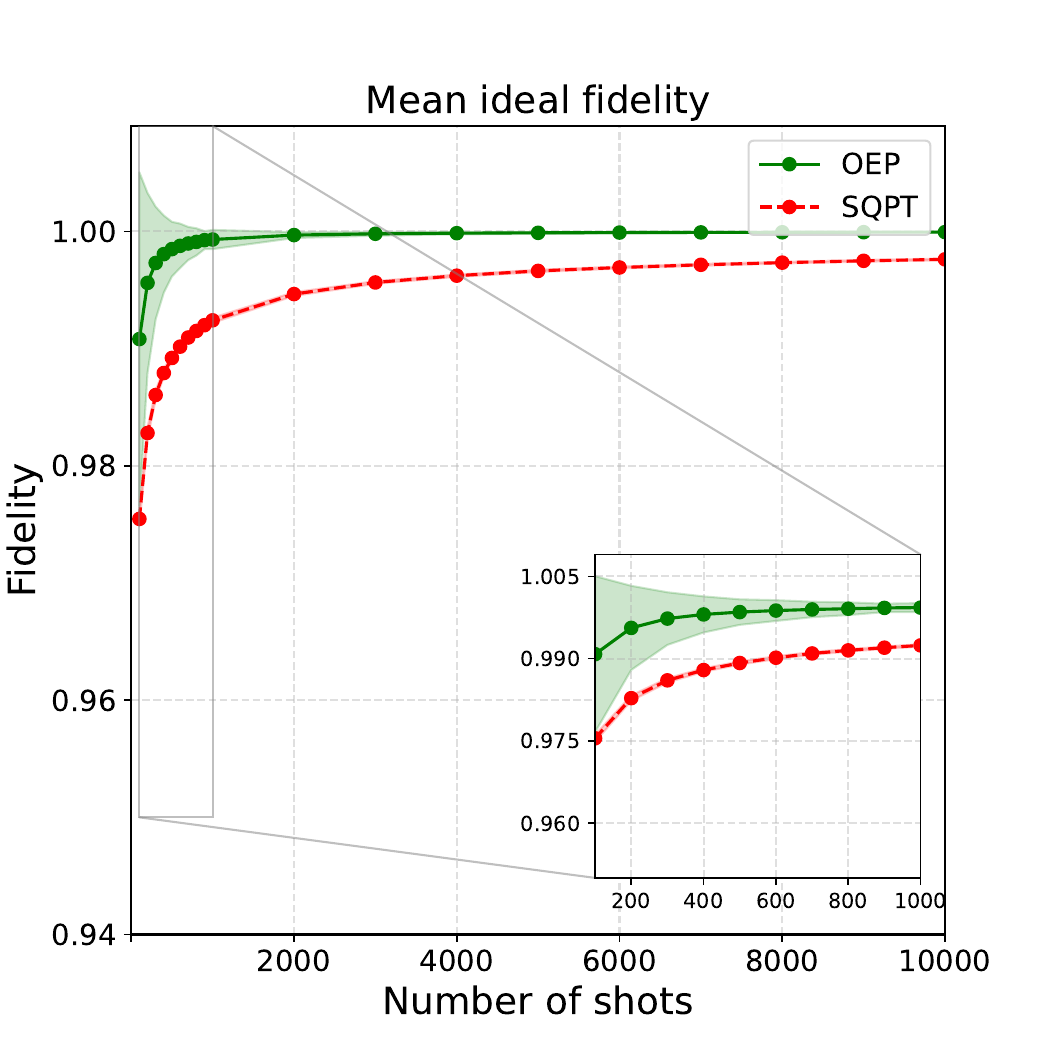} &
 \includegraphics[width=70mm]{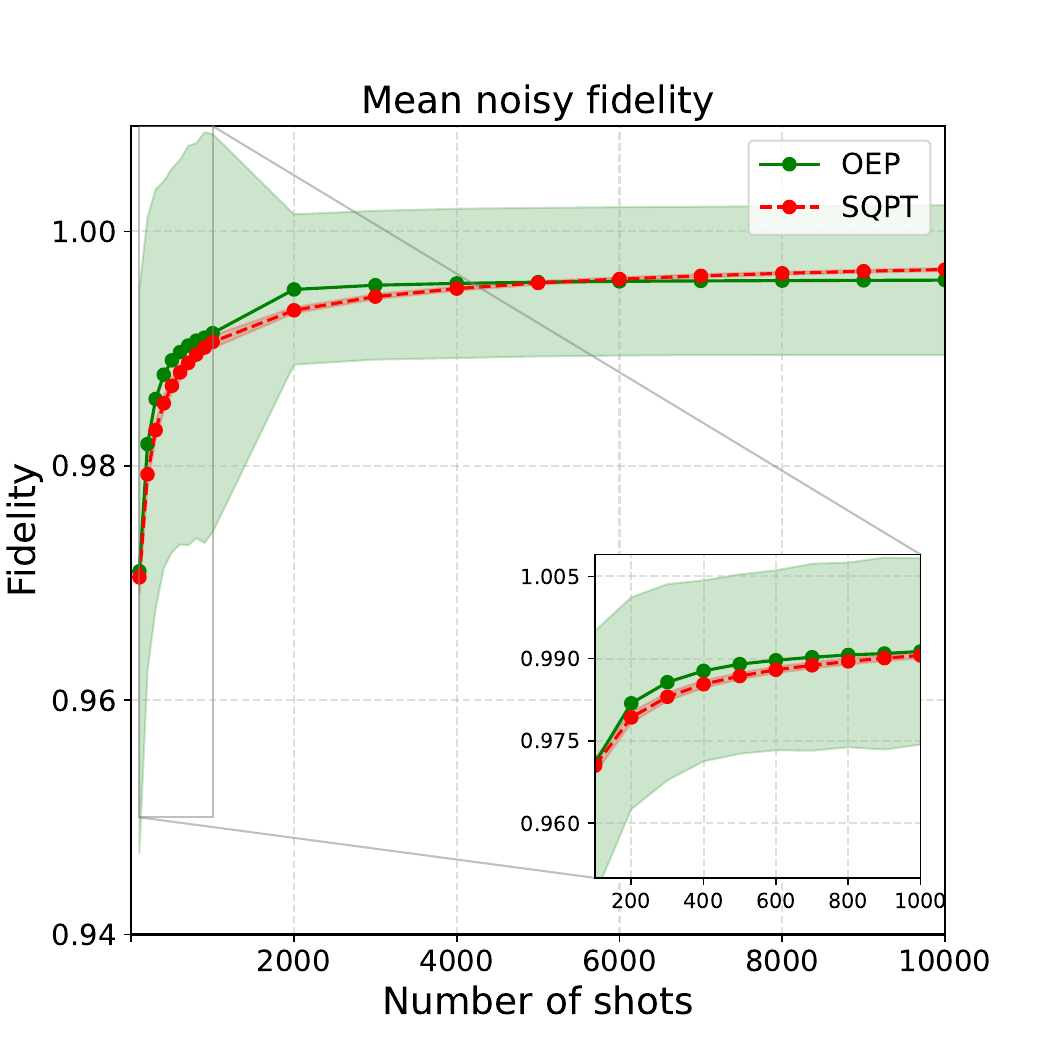}  \\
 (a) & (b) \\
 \includegraphics[width=70mm]{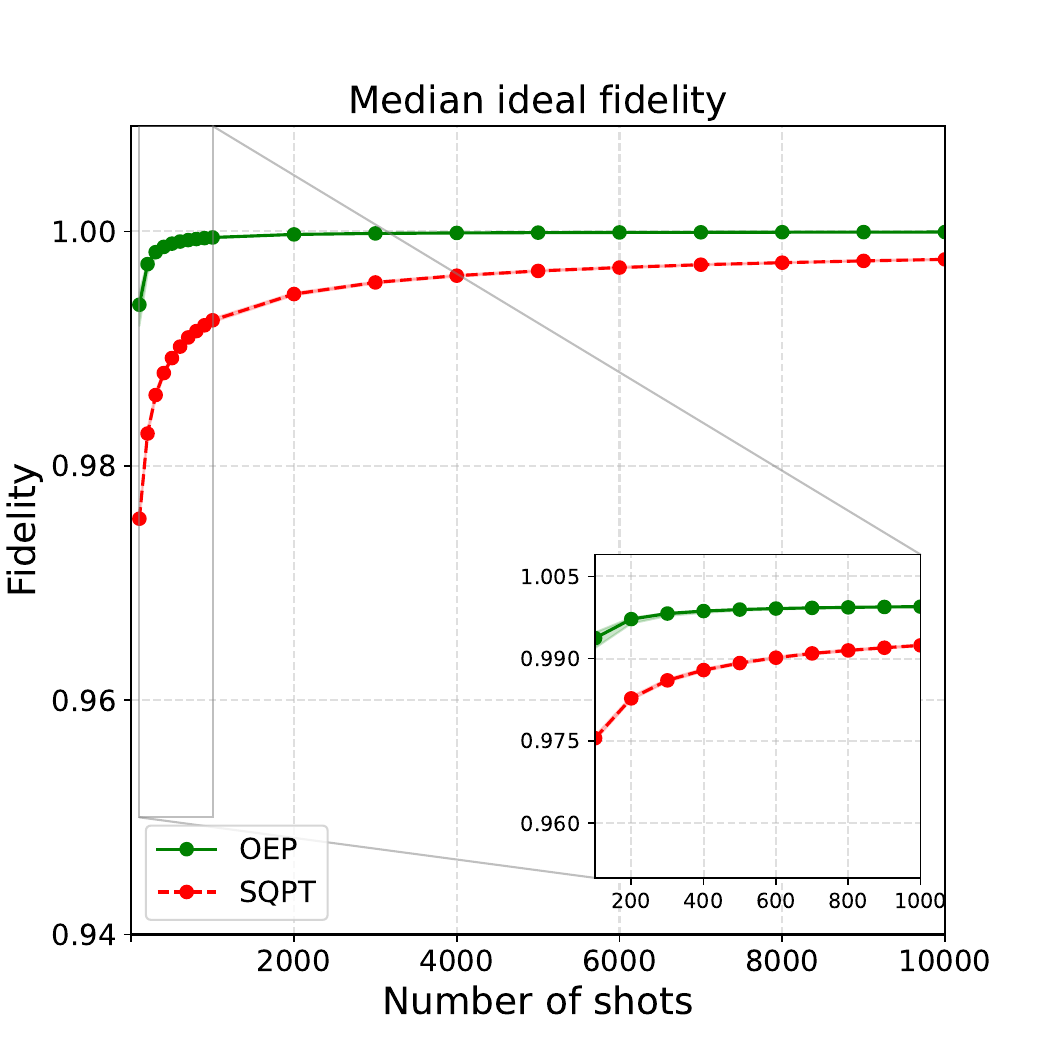} &
 \includegraphics[width=70mm]{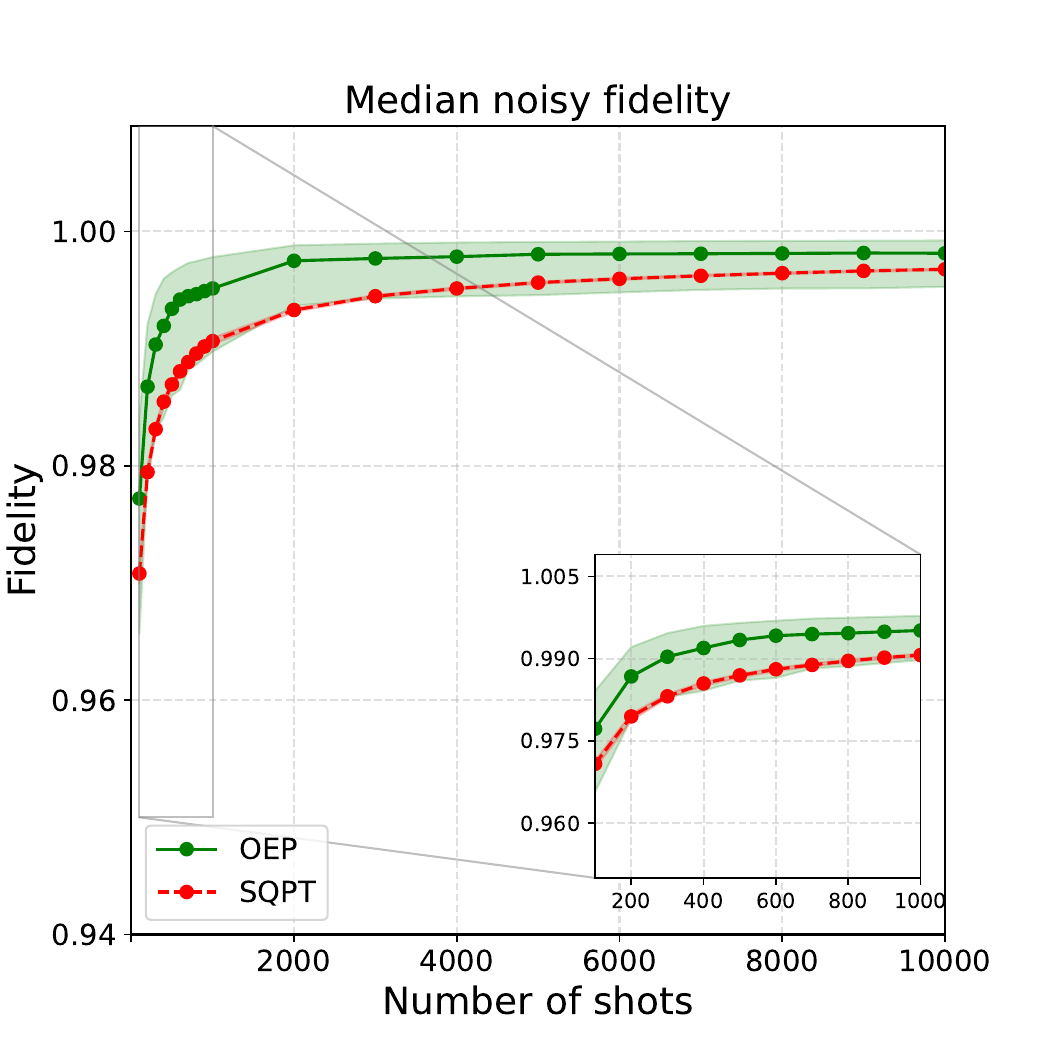} \\
 (c) & (d)
\end{tabular}
\caption{ Plots (a) and (b) show the mean and plots (c) and (d) show the median gate fidelities as functions of the number of shots for both our estimation procedure (solid green dots) and SQPT (solid red dots). The left and right plots represent the noiseless and noisy cases, respectively. Shaded areas represent standard deviation or interquartile range. The insets illustrate the low number of shots regime.}
\label{Figure1}
\end{figure*}

Figure~\ref{Figure1} shows the simulation results for our estimation procedure (green solid dots) and SQPT (red solid dots). Figures (a) and (c) show mean and median gate fidelity, respectively, as functions of the number of shots obtained in absence of error sources, that is, when the operations required by the estimation procedures are carried out perfectly. Figures (b) and (d) show results considering experiments with errors affecting single qubit gates, conditional gates, thermal relaxation, and measurements, an scenario that we call \textit{full noise model}. This situation corresponds to a unitary gate embedded in a noisy device. Insets illustrate the behavior of estimation procedures in the small number of shots regime. Shaded areas show standard deviation in Figs.~(a) and (b)  and interquartile range in Figs.~(c) and (d).

In the noiseless case, according to Figs.~(a) and (c), both our estimation procedure and SQPT are characterized by almost indistinguishable mean and median average gate fidelity. In addition, in regime of a large number of shots, both estimation procedures exhibit extremely narrow standard deviation and interquartile range. In the small number of shots regime, our estimation procedure has a very rapidly narrowing standard deviation and interquartile range. Therefore, our estimation procedure and SQPT lead to an average gate fidelity that is independent of the unitary transformation. Figures~(a) and (c) show that our estimation procedure achieves near-unit gate fidelity for ensemble sizes as small as $2\times10^3$ clearly outperforming SQPT.

Simulations show that the presence of noise affects the estimation of our procedure, decreasing the mean and median average gate fidelity with respect to their noiseless values, as exhibited in Figs.~(b) and (d). The mean average gate fidelity of both estimation procedures becomes very similar, while the median gate fidelity of our estimation procedure is above that of SQPT, although standard deviation and interquartile range of our estimation procedure become wider.

\begin{figure*}[t]
\centering
\begin{tabular}{cc}
\includegraphics[width=70mm]{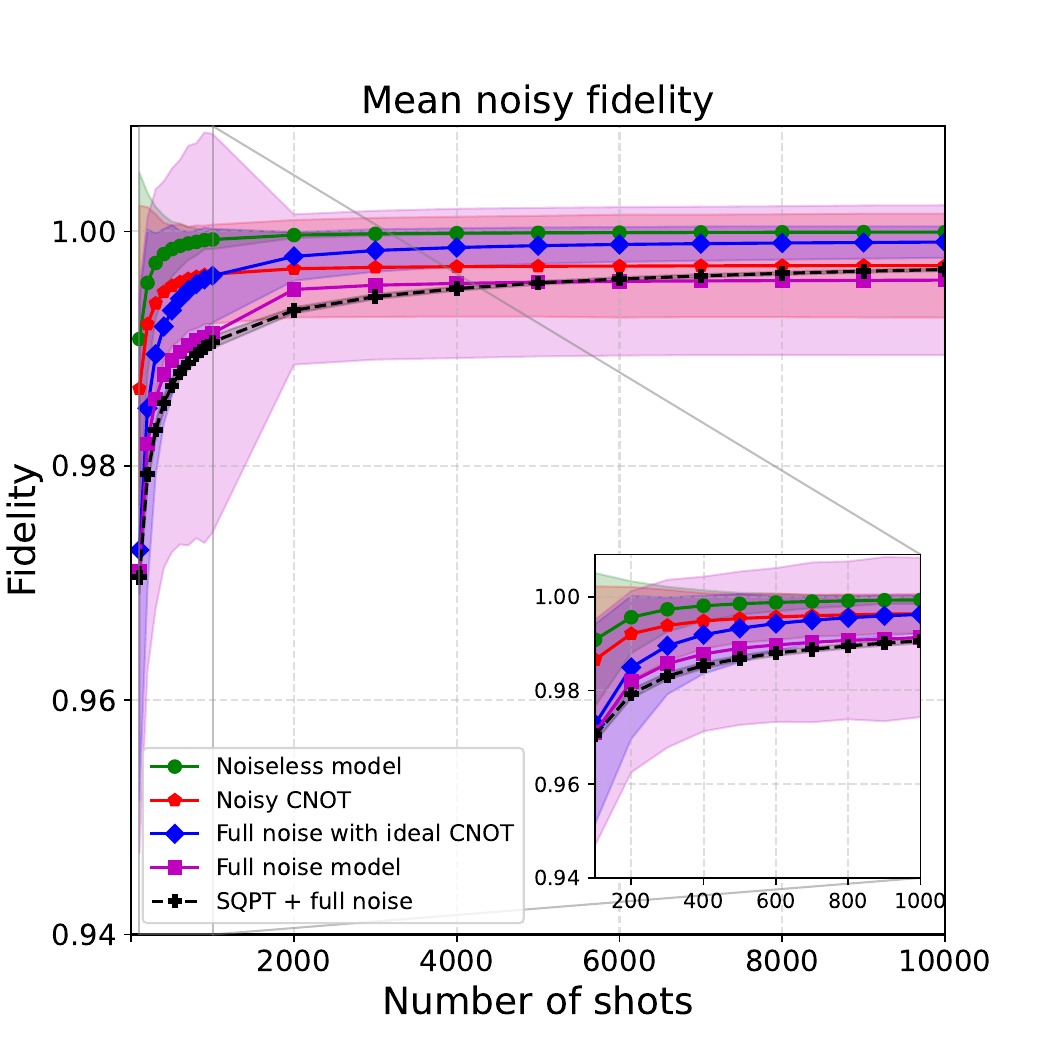} &
 \includegraphics[width=70mm]{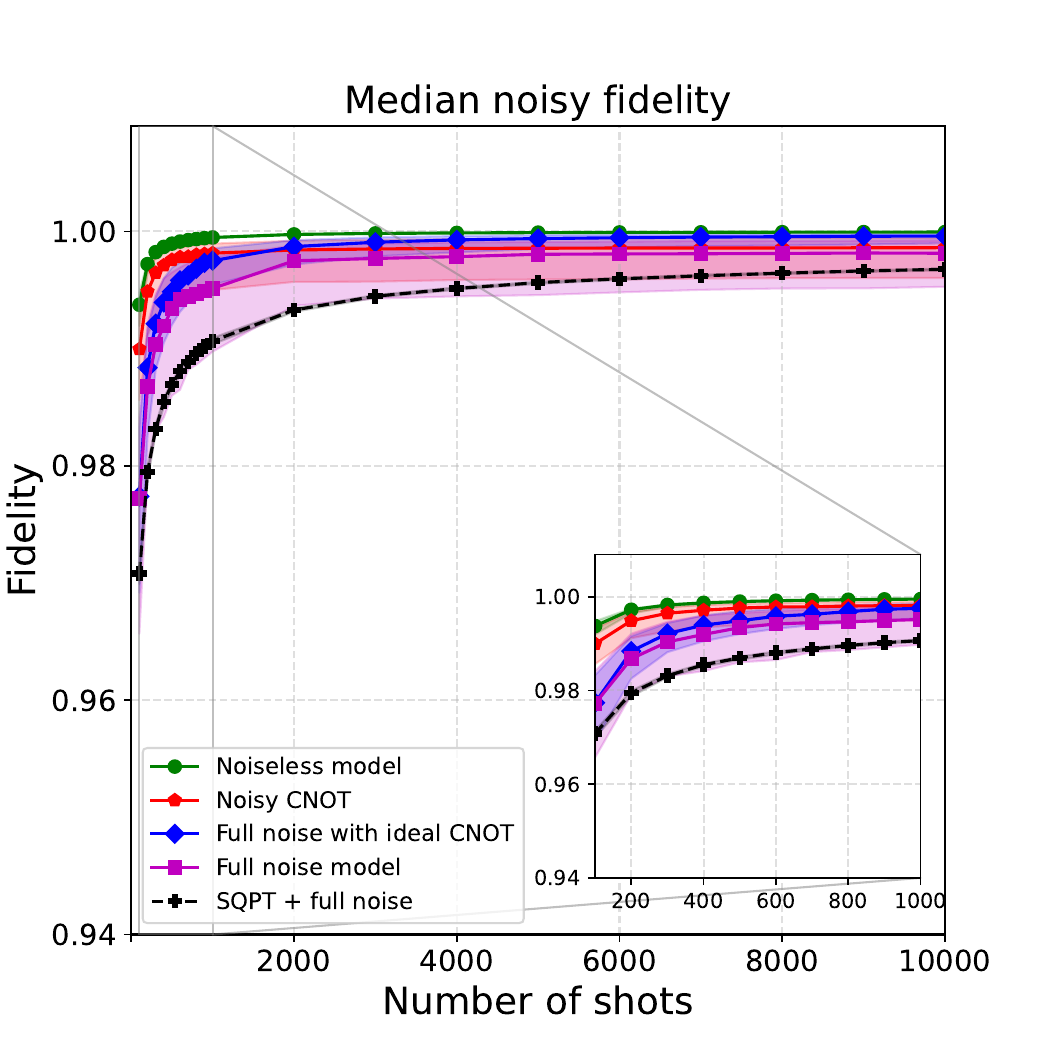}  \\
 (a) & (b) 
\end{tabular}
\caption{ Mean (a) and median (b) gate fidelities as functions of the number of shots for our estimation procedure  with different noise models and SQPT with full noise. Shaded areas represent standard deviation or interquartile range.}
\label{Figure2}
\end{figure*}

In Fig.~\ref{Figure2} we study the impact of different error sources on our estimation procedure and compare it to SQPT. Figures (a) and (b) show the mean and median average gate fidelity, respectively, for the noiseless case (solid green dots), noisy control-not gate (solid red pentagons), full noise with ideal control-not gate (solid blue diamonds) and full noise (solid pink squares) for our estimation procedure, and SQPT with full noise (black crosses). All simulations consider readout error mitigation as described in Qiskit documentation \cite{Qiskit}. The noise models and values used correspond to the \texttt{ibmq\_quito}  processor, and are provided in Appendix \ref{app:noise}. Insets depict the small ensemble regime. As expected, Fig.~\ref{Figure2} shows that the full noise model leads to the the biggest decrease in the mean and median average gate fidelity. In the small ensemble regime, the estimation considering noisy control-not gates leads to a better mean and median gate fidelity than the estimation considering other error sources. However, as the ensemble size increases, the estimation procedure is clearly more affected by noisy control-not gates. Also, in this regime mean and median average gate fidelity become constant and the increase in the ensemble has no impact in the estimation accuracy.

\section{Summary}

In this work, we propose a procedure for estimating $d$-dimensional unitary gates. Our circuit transcribes the coefficients of the gate in the Weyl-Heisenberg basis into probability amplitudes of two control qudits. For qubit gates whose Bloch vector points to a known octant, we show that our procedure saturates the quantum Cramér-Rao bound. In this case, the quantum Fisher information matrix is equivalent to the one derived in related works \cite{Ballester_2004,yuan_2016,hou_etal_2021}. We extended the analysis to higher dimensions and proved analytically that our procedure is optimal for unitary gates close to the identity in any finite dimension. In Ref.~\cite{Ballester_2004} it was shown that the quantum Cramér-Rao bound can be achieved for unitary transformations close to the identity, but no explicit protocol was proposed; our procedure accomplishes this task. Far from the identity our procedure still estimates the amplitudes $r_{p}$ of the coefficients, but the assessment of the precision of this estimation is an open problem.

In addition, we show that our estimation procedure is able to estimate any unitary transformation on qubit systems without requiring a priori information. Unitarity of the estimated operator is guaranteed by construction. Numerical simulations show that our estimation procedure outperforms SQPT in a noiseless scenario for every size of the ensemble and also in noisy scenarios with a small ensemble. Furthermore, considering a noisy scenario with ideal control-not gates, our procedure still outperforms SQPT for any ensemble size.

\section*{Acknowledgements}
This work was supported by ANID grants No. 1200266, No. 1231940, No. 1230586, No. 3230427, No. 3230407, and ANID – Millennium Science Initiative Program – ICN17$_-$012. JEM was supported by ANID-Subdireccion de Capital Humano/Doctorado Nacional/2021-21211347. 


\bibliographystyle{quantum}
\bibliography{mybibliography}


\onecolumn
\appendix
\numberwithin{equation}{section}

\section{Derivation of the output state}
\label{Final_state}

In this appendix we follow the evolution of the state through the circuit illustrated in Fig.~\ref{Circuit0}. Before that, let us start with some preliminary definitions.

Let us consider a $d$-dimensional Hilbert space $\mathcal{H}$ and the $d^2$-dimensional space $\mathcal{L}(\mathcal{H})$ of linear operators on $\mathcal{H}$. We will denote the $d^{th}$ root of unity as $\omega\equiv\exp(2\pi i/d)$, the addition modulo $d$ as $\oplus$ and the sustraction modulo $d$ as $\ominus$. Some important unitary operations in $\mathcal{L}(\mathcal{H})$ are the following:
\begin{enumerate}
    \item \textit{Identity}:
    \begin{equation}
        I:=\sum_{k=0}^{d-1}\ket{k}\bra{k}\;.
    \end{equation}
    \item \textit{Shift operator}:
    \begin{equation}
        X:=\sum_{k=0}^{d-1}\ket{k\oplus 1}\bra{k}\;.
    \end{equation}
    \item \textit{Phase operator}:
    \begin{equation}
        Z:=\sum_{k=0}^{d-1}\omega^k\ket{k}\bra{k}\;.
    \end{equation}
    \item \textit{Fourier transform}:
    \begin{equation}
        F:=\frac{1}{\sqrt{d}}\sum_{k=0}^{d-1}\omega^{jk}\ket{j}\bra{k}\;.
    \end{equation}
\end{enumerate}
It can be easily shown that the shift and phase operators satisfy the relation
\begin{equation}
\label{ZX_commuting_relation}
    Z^jX^k=\omega^{jk}X^kZ^j\;.
\end{equation}
Also, we use a short notation for the Weyl-Heisenberg operators, defined as
\begin{equation}
    D_n\equiv D_{(n_x,n_z)}:=X^{n_x}Z^{n_z}\;,
    \label{D_n}
\end{equation}
with $n=(n_x,n_z)\in \mathbb{Z}_d^2$, being $\mathbb{Z}_d=\{0,...,d-1\}$. The set $\{D_n\}_{n\in\mathbb{Z}_d^2}$ is an orthogonal basis for $\mathcal{L}(\mathcal{H})$; indeed, considering the Hilbert-Schmidt inner product for operators $\langle A,B\rangle=Tr[A^\dagger B]$ and the relation $\sum_{i=0}^{d-1}\omega^{i(j-k)}=d\delta_{jk}$, it is possible to show that
\begin{equation}
    \langle D_n,D_m\rangle=d\delta_{nm}\;,
    \label{ipWH}
\end{equation}
where $\delta_{nm}=\delta_{n_xm_x}\delta_{n_zm_z}$. Consequently, any operator in $\mathcal{L}(\mathcal{H})$ can be written as a linear combination of Weyl-Heisenberg operators. In particular, an unknown unitary $U$ can be expanded as
\begin{equation}
    U=\sum_{n\in\mathbb{Z}_d^2}r_ne^{i\phi_n}D_n=\sum_{n\in\mathbb{Z}_d^2}u_{(n_x,n_z)}X^{n_x}Z^{n_z}\;,
\end{equation}
where $u_n=r_ne^{i\phi_n}$ are complex coefficients with amplitude $r_n$ and phase $\phi_n$.

Additionally, we define the controlled unitary operators with shifted control:
\begin{equation}
V_{tc}^{(i)}:=\sum_{k=0}^{d-1}V_t^k\otimes|k\ominus i\rangle_c\langle k\ominus i|\;,
\end{equation}
where $V_t$ is a unitary gate acting on a target system $t$ controlled by a control system $c$ whose state is shifted by $i$. For our procedure, we need the following four controlled operations:
\begin{align}
    X_{02}^{(0)}&=\sum_{k=0}^{d-1}X_0^k\otimes|k\rangle_2\langle k|\;,\\
    Z_{01}^{\dagger(1)}&=\sum_{k=0}^{d-1}(Z_0^{\dagger})^k\otimes|k\ominus 1\rangle_1\langle k\ominus 1|=\sum_{k=0}^{d-1}Z_0^{-k-1}\otimes|k\rangle_1\langle k|\;,\\
    Z_{01}^{(0)}&=\sum_{k=0}^{d-1}Z_0^k\otimes|k\rangle_1\langle k|\;,\\
    X_{02}^{\dagger(1)}&=\sum_{k=0}^{d-1}(X_0^{\dagger})^k\otimes|k\ominus 1\rangle_2\langle k\ominus 1|=\sum_{k=0}^{d-1}X_0^{-k-1}\otimes|k\rangle_2\langle k|\;.
\end{align}

Now let us proceed with the protocol. We start with the initial joint state
\begin{equation}
    \ket{\Phi^0}_{012}=\ket{\psi}_0\otimes\ket{0}_1\otimes\ket{0}_2\;,
\end{equation}
where $\ket{\psi}_0$ is an arbitrary pure state. After applying the Fourier gates on the control qudits, it becomes
\begin{align}
    \ket{\Phi^1}_{012}&=(I_0\otimes F_1 \otimes F_2)\ket{\Phi^0}_{012}\\
    &=\frac{1}{d}\sum_{j_1,j_2=0}^{d-1}\ket{\psi}_0 \otimes \ket{j_1}_1 \otimes \ket{j_2}_2\;.
\end{align}
After the controlled gate $X_{02}^{(0)}$ we have
\begin{align}
    \ket{\Phi^2}_{012}&=(X_{02}^{(0)}\otimes I_1)\ket{\Phi^1}_{012}\\
    &=\frac{1}{d}\sum_{j_1,j_2=0}^{d-1}X_0^{j_2}\ket{\psi}_0\otimes\ket{j_1}_1\otimes\ket{j_2}_2\;.
\end{align}
Then,
\begin{align}
    \ket{\Phi^3}_{012}&=(Z_{01}^{\dagger(1)}\otimes I_2)\ket{\Phi^2}_{012}\\
    &=\frac{1}{d}\sum_{j_1,j_2=0}^{d-1}Z_0^{-j_1-1}X_0^{j_2}\ket{\psi}_0\otimes\ket{j_1}_1\otimes\ket{j_2}_2\;.
    \label{probe_state_app}
\end{align}
Applying $U$ on the target qudit and using Eq. \eqref{ZX_commuting_relation}, we get
\begin{align}
    \ket{\Phi^4}_{012}&=(U_0\otimes I_1\otimes I_2)\ket{\Phi^3}_{012}\\
    &=\frac{1}{d}\sum_{n_x,n_z,j_1,j_2=0}^{d-1}u_{(n_x,n_z)}\omega^{j_2(n_z-j_1-1)}X_0^{n_x+j_2}Z_0^{n_z-j_1-1}\ket{\psi}_0\otimes\ket{j_1}_1\otimes\ket{j_2}_2\;.
\end{align}
Following with the next controlled operations we have
\begin{align}
    \ket{\Phi^5}_{012}&=(Z_{01}^{(0)}\otimes I_2)\ket{\Phi^4}_{012}\\
    &=\frac{1}{d}\sum_{n_x,n_z,j_1,j_2=0}^{d-1}u_{(n_x,n_z)}\omega^{j_2(n_z-1)+j_1n_x}X_0^{n_x+j_2}Z_0^{n_z-1}\ket{\psi}_0\otimes\ket{j_1}_1\otimes\ket{j_2}_2\;,
\end{align}
and then
\begin{align}
    \ket{\Phi^6}_{012}&=(X_{02}^{\dagger(1)}\otimes I_1)\ket{\Phi^5}_{012}\\
    &=\frac{1}{d}\sum_{n_x,n_z,j_1,j_2=0}^{d-1}u_{(n_x,n_z)}\omega^{j_2(n_z-1)+j_1n_x}X_0^{n_x-1}Z_0^{n_z-1}\ket{\psi}_0\otimes\ket{j_1}_1\otimes\ket{j_2}_2\\
    &=\sum_{n_x,n_z=0}^{d-1}u_{(n_x,n_z)} X_0^{n_x-1}Z_0^{n_z-1}\ket{\psi}_0\otimes \bigg(\frac{1}{\sqrt{d}}\sum_{j_1=0}^{d-1}\omega^{j_1n_x}\ket{j_1}_1\bigg) \otimes \bigg(\frac{1}{\sqrt{d}}\sum_{j_2=0}^{d-1}\omega^{j_2(n_z-1)}\ket{j_2}_2\bigg)\\
    &=\sum_{n_x,n_z=0}^{d-1}u_{(n_x,n_z)} X_0^{n_x-1}Z_0^{n_z-1}\ket{\psi}_0\otimes F_1\ket{n_x}_1 \otimes F_2\ket{n_z\ominus1}_2\;.
\end{align}
Now we apply the inverse Fourier transform on the control qudits, getting
\begin{align}
    \ket{\Phi^7}_{012}&=(I_0\otimes F_1^{-1}\otimes F_2^{-1})\ket{\Phi^6}_{012}\\
    &=\sum_{n_x,n_z=0}^{d-1}u_{(n_x,n_z)} X_0^{n_x-1}Z_0^{n_z-1}\ket{\psi}_0\otimes \ket{n_x}_1 \otimes \ket{n_z\ominus1}_2\\
    &=\sum_{n_x,n_z=0}^{d-1}u_{(n_x,n_z\oplus1)} X_0^{n_x-1}Z_0^{n_z}\ket{\psi}_0\otimes \ket{n_x}_1 \otimes \ket{n_z}_2\;.
\end{align}
In order to disentangle the target from the control qudits, we use the following controlled gates:
\begin{align}
    X_{01}^{\dagger(-1)}&=\sum_{k=0}^{d-1}(X_0^{\dagger})^k\otimes|k\oplus 1\rangle_1\langle k\oplus 1|=\sum_{k=0}^{d-1}X_0^{-k+1}\otimes|k\rangle_1\langle k|\;,\\
    Z_{02}^{\dagger(0)}&= \sum_{k=0}^{d-1}Z_0^{-k}\otimes|k\rangle_2\langle k|\;.
\end{align}

Successive steps leads to:
\begin{align}
    \ket{\Phi^8}_{012} &= \left(X_{01}^{\dagger(1)}\otimes I_2\right)\ket{\Phi^7}_{012}\\
    &=\sum_{n_x,n_z=0}^{d-1}u_{(n_x,n_z\oplus1)} Z_0^{n_z}\ket{\psi}_0\otimes \ket{n_x}_1 \otimes \ket{n_z}_2\;,
\end{align}
\begin{align}
    \ket{\Phi^9}_{012} &= \left(Z_{02}^{\dagger(0)}\otimes I_1\right)\ket{\Phi^8}_{012}\\
    &= \ket{\psi}_0\otimes \left( \sum_{n_x,n_z=0}^{d-1}u_{(n_x,n_z\oplus1)} \ket{n_x}_1 \otimes \ket{n_z}_2\right)\;,
    \label{phi9}
\end{align}
and
\begin{align}
    \ket{\Phi^{10}}_{012} &= \left(I_0\otimes I_1 \otimes X_2\right)\ket{\Phi^9}_{012}\\
    &=\ket{\psi}_0\otimes \left( \sum_{n_x,n_z=0}^{d-1}u_{(n_x,n_z)} \ket{n_x}_1 \otimes \ket{n_z}_2\right)\\
    &=\ket{\psi}_0\otimes \ket{\varphi^0}_{12}\;,
\end{align}
where $\ket{\varphi^0}_{12} = \sum_{n_x,n_z=0}^{d-1}u_{(n_x,n_z)} \ket{n_x}_1 \otimes \ket{n_z}_2$ is the state of the control with exactly the same coefficients of $U$. For short, we write
\begin{equation}
    \ket{\varphi^0}_{12} = \sum_{n\in \mathbb{Z}_d^2}u_n\ket{n} = \sum_{n\in \mathbb{Z}_d^2}r_ne^{i\phi_n}\ket{n}\;,
    \label{varphi0}
\end{equation}
which corresponds to Eq.~\eqref{varphi0_main} in the main text.

\section{Unitarity condition}
\label{App:Unitarity}

Let us consider the unitary gate $U$, written in the Weyl-Heisenberg basis as
\begin{equation}
    U=\sum_{n\in\mathbb{Z}_d^2}r_ne^{i\phi_n}D_n\;,
\end{equation}
and its adjoint
\begin{equation}
    U^\dagger=\sum_{m\in\mathbb{Z}_d^2}r_me^{-i\phi_m}D_m^\dagger\;,
\end{equation}
where $D_n$ is defined in Eq.~\eqref{D_n}.
We have
\begin{equation}
\label{UUdag}
    UU^\dagger=\sum_{m,n\in\mathbb{Z}_d^2}r_mr_ne^{i(\phi_n-\phi_m)}D_nD_m^\dagger\;,
\end{equation}
but since $U$ is unitary we also have $UU^\dagger=I$. Noticing that $I=D_0$ and considering Eq.~\eqref{ipWH}, we have that unitarity of $U$ implies:
\begin{equation}
\label{Unitarity_conditions}
    Tr[D_p^\dagger UU^\dagger]=d\delta_{p,0}=\begin{cases}
        d & \text{if } p=(0,0) \\
        0 & \text{if } p\neq (0,0)
    \end{cases}\;.
\end{equation}
Let us calculate the left hand side of this expression using Eq. \eqref{UUdag}. We have:
\begin{align}
    Tr[D_p^\dagger UU^\dagger]&=\sum_{k=0}^{d-1} \bra{k} \sum_{m,n\in\mathbb{Z}_d^2}r_mr_ne^{i(\phi_n-\phi_m)}D_p^\dagger D_nD_m^\dagger\ket{k}\\
    &=\sum_{m,n\in\mathbb{Z}_d^2}r_mr_ne^{i(\phi_n-\phi_m)}\sum_{k=0}^{d-1} \bra{k} D_p^\dagger D_nD_m^\dagger \ket{k}\;,
\label{TrD-pUU-1}
\end{align}
but in terms of the shift and phase operators we also have
\begin{align}
    \bra{k} D_p^\dagger D_nD_m^\dagger \ket{k}&=\bra{k} (X^{p_x}Z^{p_z})^\dagger (X^{n_x}Z^{n_z})(X^{m_x}Z^{m_z})^\dagger \ket{k}\\
    &=\bra{k} Z^{-p_z}X^{-p_x}X^{n_x}Z^{n_z}Z^{-m_z}X^{-m_x} \ket{k}\\
    &=\omega^{-m_x(n_z-m_z)}\omega^{-p_z(n_x-p_x-m_x)}\bra{k} X^{n_x-p_x-m_x}Z^{n_z-m_z-p_z} \ket{k}\\
    &=\omega^{-m_x(n_z-m_z)}\omega^{-p_z(n_x-p_x-m_x)}\omega^{(n_z-m_z-p_z)k}\bra{k} X^{n_x-p_x-m_x} \ket{k}\\
    &=\omega^{-m_x(n_z-m_z)}\omega^{-p_z(n_x-p_x-m_x)}\omega^{(n_z-m_z-p_z)k}\bra{k} k\oplus n_x\ominus p_x\ominus m_x\rangle\\
    &=\omega^{-m_x(n_z-m_z)}\omega^{-p_z(n_x-p_x-m_x)}\omega^{(n_z-m_z-p_z)k}\delta_{n_x,p_x\oplus m_x}\\
    &=\omega^{-m_x(n_z-m_z)}\omega^{(n_z-m_z-p_z)k}\delta_{n_x,p_x\oplus m_x}\;,
\end{align}
where we have used the commutation relation in Eq. \eqref{ZX_commuting_relation}. Now, summing over $k$ we get
\begin{align}
    \sum_{k=0}^{d-1}\bra{k} D_p^\dagger D_nD_m^\dagger \ket{k}&=\omega^{-m_x(n_z-m_z)}\bigg(\sum_{k=0}^{d-1}\omega^{(n_z-m_z-p_z)k}\bigg)\delta_{n_x,p_x\oplus m_x}\\
    &=\omega^{-m_x(n_z-m_z)} d\delta_{n_z,p_z\oplus m_z} \delta_{n_x,p_x\oplus m_x}\\
    &=d\omega^{-m_x(n_z-m_z)} \delta_{n,p\oplus m}\;.
\label{D-pDnD-m}
\end{align}
Replacing Eq. \eqref{D-pDnD-m} in Eq. \eqref{TrD-pUU-1}, we get
\begin{align}
    Tr[D_p^\dagger UU^\dagger]&=d\sum_{m,n\in\mathbb{Z}_d^2}r_mr_ne^{i(\phi_n-\phi_m)} \omega^{-m_x(n_z-m_z)} \delta_{n,p\oplus m}\\
    &=d\sum_{m\in\mathbb{Z}_d^2}r_mr_{p\oplus m}e^{i(\phi_{p\oplus m}-\phi_m)} \omega^{-m_xp_z}\;.
\end{align}
In particular, for $p=(0,0)$ in Eq. \eqref{Unitarity_conditions} we obtain a normalization condition, which is equivalent to the total probability rule for the outcomes of our circuit:
\begin{equation}
\label{Unitary normalization}
    \sum_{m\in\mathbb{Z}_d^2}r_m^2=1\;.
\end{equation}
Besides, for $p\neq(0,0)$, we have
\begin{equation}
\label{Unitary_condition_null}
    \sum_{m\in\mathbb{Z}_d^2}r_mr_{p\oplus m}e^{i(\phi_{p\oplus m}-\phi_m)} \omega^{-m_xp_z}=0\;.
\end{equation}
Eqs. \eqref{Unitary normalization} and \eqref{Unitary_condition_null} appear as Eqs. \eqref{U-normalization} and \eqref{Unitarity-condition}, respectively, in the main text.

\section{Quantum Fisher Information Matrix}
\label{Calculation of Fisher Information matrix}

The entries $\mathcal{F}_{ab}$ of the quantum Fisher information matrix are given by
\begin{equation}
    \mathcal{F}_{ab} = 2\bra{\Phi}\{ H_a,H_b \}\ket{\Phi}- 4\bra{\Phi} H_a \ket{\Phi} \bra{\Phi} H_b \ket{\Phi}\;,
    \label{Fab}
\end{equation}
with $H_a = i\left( \partial_a U^{\dagger} \right)U = - iU^{\dagger}\left( \partial_a U\right)$, and $\ket{\Phi}$ the probe state. In our circuit, the probe state is given by Eq. \eqref{probe_state_app} as
\begin{equation}
    \ket{\Phi}=\ket{\Phi^3}_{012}=\frac{1}{d}\sum_{j_1,j_2=0}^{d-1}Z_0^{-j_1-1}X_0^{j_2}\ket{\psi}_0\otimes\ket{j_1}_1\otimes\ket{j_2}_2\;,
    \label{probe_state2}
\end{equation}
where $\ket{\psi}_0$ is the initial state of the target system.

Let us consider again the expansion of the unitary gate
\begin{equation}
    U = \sum_{n\in \mathbb{Z}_d^2} r_n e^{i\phi_n} D_{n}\;,
    \label{U_expanded}
\end{equation}
where $n = (n_x,n_z)$ is an ordered pair and $D_n = D_{n_x,n_z} = X^{n_x}Z^{n_z}$. The sum can be split using partition \eqref{partition_main} as
\begin{equation}
    U = r_0 I + \sum_{f\in S_u}r_f e^{i\phi_f} D_{f} + \sum_{a\in S_+}r_a \bigg( e^{i\phi_a} D_{a} + e^{i\phi_{\ominus a}} D_{\ominus a}\bigg)\;,
    \label{U_expanded_partition}
\end{equation}
where $r_0 = \left(1 - \sum_{n\neq (0,0)} r_n^2\right)^{1/2}$ and we have used $r_a=r_{\ominus a}$ in the last sum. Notice that
\begin{align}
    r_0&=\left(1 - \sum_{f\in S_u} r_f^2 - \sum_{a\in S_+} r_a^2 - \sum_{\ominus a\in S_-} r_{\ominus a}^2\right)^{1/2}\nonumber\\
    &=\left(1 - \sum_{f\in S_u} r_f^2 - 2\sum_{a\in S_+} r_a^2\right)^{1/2}\;.
\end{align}

We can now determine the derivatives of $U$ and therefore the operators $H_a$ that appears in Eq. \eqref{Fab}. For $f\in S_u$ we have:
\begin{equation}
    \partial_{r_f} U = - \frac{r_f}{r_0} I + e^{i\phi_f}D_f\;,
    \label{partial_rf}
\end{equation}
and therefore 
\begin{align}
    H_{r_f} &= i\left( \partial_{r_f} U^{\dagger} \right)U\\
    &= i\sum_{n\in \mathbb{Z}_d^2} r_n \left( -\dfrac{r_f}{r_0}e^{i\phi_n}D_n  + e^{i(\phi_n-\phi_f)}D_f^{\dagger}D_n\right)\;.
    \label{H_rf}
\end{align}
Similarly, for $a\in S_+$ we have:
\begin{equation}
    \partial_{r_a} U = - 2\frac{r_a}{r_0} I + e^{i\phi_a}D_a+e^{i\phi_{\ominus a}}D_{\ominus a}
    \label{partial_ra}
\end{equation}
and
\begin{equation}
    \partial_{\phi_a} U = ir_a(e^{i\phi_a}D_a-e^{i\phi_{\ominus a}}D_{\ominus a})\;,
    \label{partial_phia}
\end{equation}
leading to
\begin{equation}
    H_{r_a} = i\sum_{n\in \mathbb{Z}_d^2} r_n \left( -2\frac{r_a}{r_0}e^{i\phi_n} D_n + e^{i(\phi_n-\phi_a)}D_a^\dagger D_n + e^{i(\phi_n-\phi_{\ominus a})}D_{\ominus a}^\dagger D_n\right)\;
    \label{H_ra}
\end{equation}
and
\begin{equation}
    H_{\phi_a} = \sum_{n\in \mathbb{Z}_d^2} r_ar_n \left( e^{i(\phi_n-\phi_a)}D_a^\dagger D_n -e^{i(\phi_n-\phi_{\ominus a})}D_{\ominus a}^\dagger D_n\right)\;.
    \label{H_phia}
\end{equation}

In order to calculate the anti-commutators in Eq. \eqref{Fab}, let us note that 
\begin{align}
    \{ H_a , H_b \} &= \left[ i\left( \partial_a U^{\dagger} \right)U \right]\left[ -iU^{\dagger}\left( \partial_b U\right)\right] + \left[ i\left( \partial_b U^{\dagger} \right)U \right]\left[ -iU^{\dagger}\left( \partial_a U\right)\right] \nonumber \\
    &=\left( \partial_a U^{\dagger} \right)UU^{\dagger}\left( \partial_b U\right) + \left( \partial_b U^{\dagger} \right)UU^{\dagger}\left( \partial_a U\right)\;.
    \label{HaHb1}
\end{align}
By imposing unitarity on $U$ we have that $UU^{\dagger}=I$ and Eq. \eqref{HaHb1} becomes
\begin{equation}
    \{ H_a , H_b \} = \left( \partial_a U^{\dagger} \right) \left( \partial_b U\right) + \left( \partial_b U^{\dagger} \right)\left( \partial_a U\right)\,.
    \label{HaHb}
\end{equation}

We can now replace Eqs. \eqref{partial_rf}, \eqref{partial_ra} and \eqref{partial_phia} into \eqref{HaHb} and get:
\begin{align}
    \{ H_{r_f} , H_{r_g} \} =&\;  2\frac{r_f r_g}{r_0^2} I - \dfrac{r_f}{r_0}\left( e^{i\phi_g}D_g + e^{-i\phi_g}D_g^{\dagger}  \right)- \dfrac{r_g}{r_0}\left( e^{i\phi_f}D_f + e^{-i\phi_f}D_f^{\dagger}  \right)\nonumber \\
    &+ e^{i(\phi_g - \phi_f)}D_f^{\dagger}D_g + e^{i(\phi_f - \phi_g)}D_g^{\dagger}D_f 
    \label{H_rf_H_rg}\;,\\\nonumber\\
    \{ H_{r_f} , H_{r_a} \} =&\; 4\frac{r_fr_a}{r_0^2}I - 2\frac{r_a}{r_0}\left( e^{i\phi_f}D_f+e^{-i\phi_f}D_f^\dagger\right) - \frac{r_f}{r_0}\left( e^{i\phi_a}D_a+e^{-i\phi_a}D_a^\dagger + e^{i\phi_{\ominus a}}D_{\ominus a}+e^{-i\phi_{\ominus a}}D_{\ominus a}^\dagger \right) \nonumber\\
    & + \left( e^{i(\phi_a - \phi_f)}D_f^{\dagger}D_a + e^{i(\phi_f - \phi_a)}D_a^{\dagger}D_f\right) + \left( e^{i(\phi_{\ominus a} - \phi_f)}D_f^{\dagger}D_{\ominus a} + e^{i(\phi_f - \phi_{\ominus a})}D_{\ominus a}^{\dagger}D_f\right) \;,
    \label{H_rf_H_ra}\\
    \{ H_{r_f} , H_{\phi_a} \} =&\; ir_a\bigg( \frac{r_f}{r_0} \left( -e^{i\phi_a}D_a + e^{-i\phi_a}D_a^\dagger + e^{i\phi_{\ominus a}}D_{\ominus a} - e^{-i\phi_{\ominus a}}D_{\ominus a}^\dagger\right) \nonumber\\
    &+ \left( e^{i(\phi_a - \phi_f)}D_f^{\dagger}D_a - e^{i(\phi_f - \phi_a)}D_a^{\dagger}D_f\right) + \left( - e^{i(\phi_{\ominus a} - \phi_f)}D_f^{\dagger}D_{\ominus a} + e^{i(\phi_f - \phi_{\ominus a})}D_{\ominus a}^{\dagger}D_f\right) \bigg) \;,
    \label{H_rf_H_phia}\\
    \{ H_{r_a} , H_{r_b} \} =&\; 8\frac{r_ar_b}{r_0^2}I - 2\frac{r_a}{r_0} \left( e^{i\phi_b}D_b + e^{-i\phi_b}D_b^\dagger + e^{i\phi_{\ominus b}}D_{\ominus b} + e^{-i\phi_{\ominus b}}D_{\ominus b}^\dagger \right) \nonumber\\
    & - 2\frac{r_b}{r_0} \left( e^{i\phi_a}D_a + e^{-i\phi_a}D_a^\dagger + e^{i\phi_{\ominus a}}D_{\ominus a} + e^{-i\phi_{\ominus a}}D_{\ominus a}^\dagger \right) \nonumber\\
    & + \left( e^{i(\phi_b-\phi_a)}D_a^\dagger D_b + e^{i(\phi_a-\phi_b)}D_b^\dagger D_a \right) + \left( e^{i(\phi_{\ominus b}-\phi_a)}D_a^\dagger D_{\ominus b} + e^{i(\phi_a-\phi_{\ominus b})}D_{\ominus b}^\dagger D_a \right) \nonumber\\
    & + \left( e^{i(\phi_b-\phi_{\ominus a})}D_{\ominus a}^\dagger D_b + e^{i(\phi_{\ominus a}-\phi_b)}D_b^\dagger D_{\ominus a} \right) + \left( e^{i(\phi_{\ominus b}-\phi_{\ominus a})}D_{\ominus a}^\dagger D_{\ominus b} + e^{i(\phi_{\ominus a}-\phi_{\ominus b})}D_{\ominus b}^\dagger D_{\ominus a} \right)\;,
    \label{H_ra_H_rb}
\end{align}
\begin{align}
    \{ H_{r_a} , H_{\phi_b} \} =&\; 2i\frac{r_ar_b}{r_0}\left( -e^{i\phi_b}D_b + e^{-i\phi_b}D_b^\dagger + e^{i\phi_{\ominus b}}D_{\ominus b} - e^{-i\phi_{\ominus b}}D_{\ominus b}^\dagger \right) \nonumber\\
    & + i r_b\bigg( e^{i(\phi_b-\phi_a)}D_a^\dagger D_b - e^{i(\phi_a-\phi_b)}D_b^\dagger D_a - e^{i(\phi_{\ominus b}-\phi_a)}D_a^\dagger D_{\ominus b} + e^{i(\phi_a-\phi_{\ominus b})}D_{\ominus b}^\dagger D_a \nonumber\\
    & + e^{i(\phi_b-\phi_{\ominus a})}D_{\ominus a}^\dagger D_b - e^{i(\phi_{\ominus a}-\phi_b)}D_b^\dagger D_{\ominus a} + e^{i(\phi_{\ominus a}-\phi_{\ominus b})}D_{\ominus b}^\dagger D_{\ominus a} - e^{i(\phi_{\ominus b}-\phi_{\ominus a})}D_{\ominus a}^\dagger D_{\ominus b} \bigg)
    \label{H_ra_H_phib}
\end{align}
and
\begin{align}
    \{ H_{\phi_a} , H_{\phi_b} \} =&\; r_ar_b \bigg( e^{i(\phi_b-\phi_a)}D_a^\dagger D_b + e^{i(\phi_a-\phi_b)}D_b^\dagger D_a - e^{i(\phi_{\ominus b}-\phi_a)}D_a^\dagger D_{\ominus b} - e^{i(\phi_a-\phi_{\ominus b})}D_{\ominus b}^\dagger D_a \nonumber\\
    & - e^{i(\phi_b-\phi_{\ominus a})}D_{\ominus a}^\dagger D_b - e^{i(\phi_{\ominus a}-\phi_b)}D_b^\dagger D_{\ominus a} + e^{i(\phi_{\ominus a}-\phi_{\ominus b})}D_{\ominus b}^\dagger D_{\ominus a} + e^{i(\phi_{\ominus b}-\phi_{\ominus a})}D_{\ominus a}^\dagger D_{\ominus b} \bigg)\;.
    \label{H_phia_H_phib}
\end{align}

In order to calculate the expectation values $\bra{\Phi}H_a\ket{\Phi}$ and $\bra{\Phi}\{ H_a,H_b \}\ket{\Phi}$ in Eq. \eqref{Fab}, we are going to need $\bra{\Phi}D_n\ket{\Phi}$ and $\bra{\Phi}D_a^\dagger D_b\ket{\Phi}$. We can simplify the notation for our probe state and rewrite Eq. \eqref{probe_state2} as 
\begin{equation}
    \ket{\Phi} = \frac{1}{d}\sum_{r,s = 0}^{d - 1} Z^{-1-r}X^{s}\ket{\psi}\ket{rs}\;.
\end{equation}
Thus,
\begin{align}
    \bra{\Phi}D_n\ket{\Phi} &= \frac{1}{d^2}\sum_{r',s'=0}^{d-1}\sum_{r,s=0}^{d-1}\bra{\psi}\bra{r's'} X^{-s'}Z^{1+r'}D_n Z^{-1-r}X^{s}\ket{\psi}\ket{rs} \nonumber\\
    &= \frac{1}{d^2}\sum_{r',s'=0}^{d-1}\sum_{r,s=0}^{d-1}\bra{r's'}rs\rangle \bra{\psi}X^{-s'}Z^{1+r'}D_n Z^{-1-r}X^{s}\ket{\psi} \nonumber\\
    &= \frac{1}{d^2}\sum_{r',s'=0}^{d-1}\sum_{r,s=0}^{d-1}\delta_{r'r}\delta_{s's} \bra{\psi}X^{-s'}Z^{1+r'}D_n Z^{-1-r}X^{s}\ket{\psi} \nonumber\\
    &= \frac{1}{d^2}\sum_{r,s}\bra{\psi}X^{-s}Z^{1+r}D_n Z^{-1-r}X^{s}\ket{\psi}\;.
\end{align}
But $X^{-s}Z^{1+r}D_n Z^{-1-r}X^{s} = X^{-s}Z^{1+r}(X^{n_x}Z^{n_z}) Z^{-1-r}X^{s}=X^{n_x}Z^{n_z}\omega^{r n_x}\omega^{s n_z}\omega^{n_x}$. Thus,
\begin{align}
    \bra{\Phi}D_n\ket{\Phi} &=  \frac{1}{d^2}\bra{\psi}X^{n_x}Z^{n_z}\ket{\psi}\omega^{n_x}\sum_{r}\omega^{r n_x}\sum_{s}\omega^{s n_z} \nonumber \\
    &= \frac{1}{d^2}\bra{\psi}D_n\ket{\psi}\omega^{n_x}d^2\delta_{n_x,0}\delta_{n_z,0} \nonumber\\
    &= \bra{\psi}D_n\ket{\psi}\omega^{n_x}\delta_{n_x,0}\delta_{n_z,0}\;.
    \label{exp_Dn1}
\end{align}
Note that Eq. \eqref{exp_Dn1} is different to zero only when $n_x = 0$ and $n_z=0$, in which case $D_{n} = D_0 = I$. Hence,
\begin{equation}
    \bra{\Phi}D_n\ket{\Phi}=\delta_{n,0}\;.
    \label{exp_Dn}
\end{equation}

Also,
\begin{align}
    D_a^{\dagger}D_b  &= Z^{-a_z}X^{-a_x}X^{b_x}Z^{b_z} \nonumber\\
    &= Z^{-a_z}X^{b_x-a_x}Z^{b_z} \nonumber\\
    &= \omega^{-a_z(b_x - a_x)} X^{b_x-a_x}Z^{b_z-a_z} \nonumber\\
    &= \omega^{-a_z(b_x - a_x)} D_{b-a}\;.
\end{align}
Thus, $\bra{\Phi}D_a^{\dagger}D_b\ket{\Phi} = \omega^{-a_z(b_x - a_x)}\bra{\Phi}D_{b-a}\ket{\Phi}$. Replacing $D_{b-a}$ into \eqref{exp_Dn}  we obtain 
\begin{align}
    \bra{\Phi}D_a^{\dagger}D_b\ket{\Phi} &= \omega^{-a_z(b_x - a_x)}\bra{\Phi}D_{b-a}\ket{\Phi} \nonumber\\
    &= \omega^{-a_z(b_x - a_x)}\delta_{a_x,b_x}\delta_{a_z,b_z} \nonumber\\
    &= \delta_{a,b}
    \label{DaDb}
\end{align}

We can now find each $\bra{\Phi}H_a\ket{\Phi}$. By using Eq. \eqref{H_rf},
\begin{equation}
    \label{exp_Hf}
    \bra{\Phi}H_{r_f}\ket{\Phi} = i \sum_n\left( -\dfrac{r_f r_n}{r_0}e^{i\phi_n}\bra{\Phi}D_n\ket{\Phi} + r_n e^{i(\phi_n-\phi_f)}\bra{\Phi}D_f^{\dagger}D_n\ket{\Phi} \right)\;.
\end{equation}
From Eqs.\eqref{exp_Dn} and \eqref{DaDb} we see that the first term in Eq.\eqref{exp_Hf} is non-zero only for $n=0$, whereas the second term is non-zero only for $n=f$. Therefore,
\begin{equation}
    \bra{\Phi}H_{r_f}\ket{\Phi} = -i \frac{r_f r_0}{r_0} + i r_f = 0\;.
    \label{expect_H_rf}
\end{equation}
Similarly, using Eq. \eqref{H_ra} we have
\begin{align}
    \bra{\Phi}H_{r_a}\ket{\Phi} &= i\sum_{n\in \mathbb{Z}_d^2} r_n \left( -2\frac{r_a}{r_0}e^{i\phi_n} \bra{\Phi}D_n\ket{\Phi} + e^{i(\phi_n-\phi_a)}\bra{\Phi}D_a^\dagger D_n \ket{\Phi}+ e^{i(\phi_n-\phi_{\ominus a})}\bra{\Phi}D_{\ominus a}^\dagger D_n\ket{\Phi}\right)\nonumber\\
    &= i\left( -2\frac{r_ar_0}{r_0} + r_a +r_{\ominus a}\right)\nonumber\\
    &= 0\;,
    \label{expect_H_ra}
\end{align}
and using Eq. \eqref{H_phia} we get
\begin{align}
    \bra{\Phi}H_{\phi_a}\ket{\Phi} &= \sum_{n\in \mathbb{Z}_d^2} r_ar_n \left( e^{i(\phi_n-\phi_a)}\bra{\Phi}D_a^\dagger D_n\ket{\Phi} -e^{i(\phi_n-\phi_{\ominus a})}\bra{\Phi}D_{\ominus a}^\dagger D_n\ket{\Phi}\right)\nonumber\\
    &= r_a^2 - r_ar_{\ominus a} \nonumber\\
    &= 0\;.
    \label{expect_H_phia}
\end{align}

Now let us determine the expectation values of the form $\bra{\Phi}\{ H_a, H_b \}\ket{\Phi}$. From \eqref{H_rf_H_rg} and using Eqs.\eqref{exp_Dn} and \eqref{DaDb} we get
\begin{equation}
    \bra{\Phi}\{ H_{r_f}, H_{r_g} \}\ket{\Phi} = 2\frac{r_fr_g}{r_0^2}-\frac{r_f}{r_0}(e^{i\phi_g}+e^{-i\phi_g})\delta_{g,0}-\frac{r_g}{r_0}(e^{i\phi_f}+e^{-i\phi_f})\delta_{f,0}+(e^{i(\phi_g-\phi_f)}+e^{i(\phi_f-\phi_g)})\delta_{f,g}\;.
    \label{exp_HfHf_1}
\end{equation}
Given the partition \eqref{partition_main}, we have $f,g\in S_u$. It follows that $f\neq 0$ and $g\neq 0$, hence the second and third term vanish. Thus,
\begin{equation}
    \bra{\Phi}\{ H_{r_f}, H_{r_g} \}\ket{\Phi} = 2\frac{r_fr_g}{r_0^2}+2\delta_{f,g}\;.
    \label{expect_H_rf_H_rg}
\end{equation}
By replacing Eqs. \eqref{expect_H_rf} and \eqref{expect_H_rf_H_rg} into \eqref{Fab}, we obtain the entries of the first block of $\mathcal{F}$:
\begin{equation}
    \mathbf{A}_{f,g}:=\mathcal{F}_{r_fr_g} = 4\frac{r_fr_g}{r_0^2} + 4\delta_{f,g}\;.
    \label{F_rf_rg}
\end{equation}
In the same way, from Eq. \eqref{H_rf_H_ra}, we have
\begin{align}
    \bra{\Phi}\{ H_{r_f}, H_{r_a} \}\ket{\Phi} = &\; 4\frac{r_fr_a}{r_0^2} - 2\frac{r_a}{r_0}\left( e^{i\phi_f}+e^{-i\phi_f}\right)\delta_{f,0} \nonumber\\
    & - \frac{r_f}{r_0}\left( e^{i\phi_a}\delta_{a,0}+e^{-i\phi_a}\delta_{a,0} + e^{i\phi_{\ominus a}}\delta_{\ominus a,0}+e^{-i\phi_{\ominus a}}\delta_{\ominus a,0} \right) \nonumber\\
    & + \left( e^{i(\phi_a - \phi_f)} + e^{i(\phi_f - \phi_a)}\right) \delta_{f,a} + \left( e^{i(\phi_{\ominus a} - \phi_f)} + e^{i(\phi_f - \phi_{\ominus a})}\right)\delta_{f,\ominus a}\;.
\end{align}
Notice that all the Kronecker deltas in this equation are null because their indexes belong to different sets of the partition \eqref{partition_main}. Thus,
\begin{equation}
    \bra{\Phi}\{ H_{r_f}, H_{r_a} \}\ket{\Phi} =  4\frac{r_fr_a}{r_0^2}\;,
\end{equation}
and we get the second block of $\mathcal{F}$:
\begin{equation}
    \mathbf{B}_{f,a}:=\mathcal{F}_{r_fr_a} = 8\frac{r_fr_a}{r_0^2}\;.
    \label{F_rf_ra}
\end{equation}

From Eq. \eqref{H_rf_H_phia}, we have
\begin{align}
    \bra{\Phi}\{ H_{r_f}, H_{\phi_a} \}\ket{\Phi} =&\; ir_a\bigg( \frac{r_f}{r_0} \left( -e^{i\phi_a}\delta_{a,0} + e^{-i\phi_a}\delta_{a,0} + e^{i\phi_{\ominus a}}\delta_{\ominus a,0} - e^{-i\phi_{\ominus a}}\delta_{\ominus a,0}\right)\nonumber\\
    &+ \left( e^{i(\phi_a - \phi_f)} - e^{i(\phi_f - \phi_a)}\right)\delta_{f,a} + \left( - e^{i(\phi_{\ominus a} - \phi_f)} + e^{i(\phi_f - \phi_{\ominus a})}\right)\delta_{f,\ominus a} \bigg) \;.
\end{align}
Again, the Kronecker deltas are always zero because of the partition, hence
\begin{equation}
    \bra{\Phi}\{ H_{r_f}, H_{\phi_a} \}\ket{\Phi}=0\;,
\end{equation}
and thus
\begin{equation}
    \mathcal{F}_{r_f\phi_a} =0\;.
    \label{F_rf_phia}
\end{equation}

From Eq. \eqref{H_ra_H_rb} we have
\begin{align}
    \bra{\Phi}\{ H_{r_a}, H_{r_b} \}\ket{\Phi} = &\; 8\frac{r_ar_b}{r_0^2} - 2\frac{r_a}{r_0} \left( e^{i\phi_b}\delta_{b,0} + e^{-i\phi_b}\delta_{b,0} + e^{i\phi_{\ominus b}}\delta_{\ominus b,0} + e^{-i\phi_{\ominus b}}\delta_{\ominus b,0} \right) \nonumber\\
    & - 2\frac{r_b}{r_0} \left( e^{i\phi_a}\delta_{a,0} + e^{-i\phi_a}\delta_{a,0} + e^{i\phi_{\ominus a}}\delta_{\ominus a,0} + e^{-i\phi_{\ominus a}}\delta_{\ominus a,0} \right) \nonumber\\
    & + \left( e^{i(\phi_b-\phi_a)} + e^{i(\phi_a-\phi_b)} \right)\delta_{a,b} + \left( e^{i(\phi_{\ominus b}-\phi_a)} + e^{i(\phi_a-\phi_{\ominus b})} \right)\delta_{a,\ominus b} \nonumber\\
    & + \left( e^{i(\phi_b-\phi_{\ominus a})} + e^{i(\phi_{\ominus a}-\phi_b)} \right)\delta_{\ominus a,b} + \left( e^{i(\phi_{\ominus b}-\phi_{\ominus a})} + e^{i(\phi_{\ominus a}-\phi_{\ominus b})} \right)\delta_{a,b}\;,
\end{align}
Now, the only Kronecker deltas that do not vanish are $\delta_{a,b}$. Thus,
\begin{equation}
    \bra{\Phi}\{ H_{r_a}, H_{r_b} \}\ket{\Phi} = 8\frac{r_ar_b}{r_0^2} + 4\delta_{a,b}
\end{equation}
and
\begin{equation}
    \mathbf{C}_{a,b}:=\mathcal{F}_{r_ar_b} = 16\frac{r_ar_b}{r_0^2} + 8\delta_{a,b}\;.
    \label{F_ra_rb}
\end{equation}

From Eq. \eqref{H_ra_H_phib},
\begin{align}
    \bra{\Phi}\{ H_{r_a}, H_{\phi_b} \}\ket{\Phi} = &\; 2i\frac{r_ar_b}{r_0}\left( -e^{i\phi_b}\delta_{b,0} + e^{-i\phi_b}\delta_{b,0} + e^{i\phi_{\ominus b}}\delta_{\ominus b,0} - e^{-i\phi_{\ominus b}}\delta_{\ominus b,0} \right) \nonumber\\
    & + i r_b\bigg( e^{i(\phi_b-\phi_a)}\delta_{a,b} - e^{i(\phi_a-\phi_b)}\delta_{a,b} - e^{i(\phi_{\ominus b}-\phi_a)}\delta_{a,\ominus b} + e^{i(\phi_a-\phi_{\ominus b})}\delta_{a,\ominus b} \nonumber\\
    & + e^{i(\phi_b-\phi_{\ominus a})}\delta_{\ominus a,b} - e^{i(\phi_{\ominus a}-\phi_b)}\delta_{\ominus a,b} + e^{i(\phi_{\ominus a}-\phi_{\ominus b})}\delta_{a,b} - e^{i(\phi_{\ominus b}-\phi_{\ominus a})}\delta_{a,b} \bigg)\;,
\end{align}
but some Kronecker deltas are null because of the partition, while the terms in $\delta_{a,b}$ cancel out. Thus,
\begin{equation}
    \bra{\Phi}\{ H_{r_a}, H_{\phi_b} \}\ket{\Phi}=0\;
\end{equation}
and
\begin{equation}
    \mathcal{F}_{r_a\phi_b} =0\;.
    \label{F_ra_phib}
\end{equation}

Finally, from Eq. \eqref{H_phia_H_phib}, we have
\begin{align}
    \bra{\Phi}\{ H_{\phi_a}, H_{\phi_b} \}\ket{\Phi} = &\; r_ar_b \bigg( e^{i(\phi_b-\phi_a)}\delta_{a,b} + e^{i(\phi_a-\phi_b)}\delta_{a,b} - e^{i(\phi_{\ominus b}-\phi_a)}\delta_{a,\ominus b} - e^{i(\phi_a-\phi_{\ominus b})}\delta_{a,\ominus b} \nonumber\\
    & - e^{i(\phi_b-\phi_{\ominus a})}\delta_{\ominus a,b} - e^{i(\phi_{\ominus a}-\phi_b)}\delta_{\ominus a,b} + e^{i(\phi_{\ominus a}-\phi_{\ominus b})}\delta_{a,b} + e^{i(\phi_{\ominus b}-\phi_{\ominus a})}\delta_{a,b} \bigg)\nonumber\\
    = &\; 4r_ar_b\delta_{a,b}\nonumber\\
    = &\; 4r_a^2 \delta_{a,b}
\end{align}
and
\begin{equation}
    \mathbf{D}_{a,b}:=\mathcal{F}_{\phi_a\phi_b} = 8r_a^2 \delta_{a,b}\;.
    \label{F_phia_phib}
\end{equation}

Summarizing, the quantum Fisher information matrix for estimating a unitary gate close-to-the-identity in dimension even is
\begin{equation}
    \mathcal{F}_{even} = \begin{pmatrix}
        \mathbf{A} & \mathbf{B} & \mathbf{0}\\
        \mathbf{B}^T & \mathbf{C} & \mathbf{0}\\
        \mathbf{0} & \mathbf{0} & \mathbf{D}
    \end{pmatrix}\;,
    \label{F_even}
\end{equation}
with $\mathbf{A}$, $\mathbf{B}$, $\mathbf{C}$, and $\mathbf{D}$ being the blocks defined in Eqs. \eqref{F_rf_rg}, \eqref{F_rf_ra}, \eqref{F_ra_rb} and \eqref{F_phia_phib}, respectively. This matrix corresponds to the block matrix in Eq. \eqref{F_even_main}.

\section{Classical Fisher Information Matrix}
\label{d_dimensional_Classic_Fisher_Appendix}

The classical Fisher information matrix is given by 
\begin{equation}
    \mathcal{I}_{ab}=\sum_y \frac{1}{P(y|\mathbf{t})}\left[\frac{\partial P(y|\mathbf{t})}{\partial t_a}\right]\left[\frac{\partial P(y|\mathbf{t})}{\partial t_b}\right]\;.
\end{equation}
We can expand the sum using the partition \eqref{partition_main}:
\begin{equation}
    \mathcal{I}_{ab}=\frac{1}{P_0}\left[\frac{\partial P_0}{\partial t_a}\right]\left[\frac{\partial P_0}{\partial t_b}\right] + \sum_{h\in S_u} \frac{1}{P_h}\left[\frac{\partial P_h}{\partial t_a}\right]\left[\frac{\partial P_h}{\partial t_b}\right] + \sum_{c\in S_+} \left(\frac{1}{P_c}\left[\frac{\partial P_c}{\partial t_a}\right]\left[\frac{\partial P_c}{\partial t_b}\right] + \frac{1}{P_{\ominus c}}\left[\frac{\partial P_{\ominus c}}{\partial t_a}\right]\left[\frac{\partial P_{\ominus c}}{\partial t_b}\right]\right)\;.
    \label{I_expanded}
\end{equation}

Deriving the probablities in Eqs. \eqref{probs_main} we get:
\begin{align}
    \frac{\partial P_0}{\partial r_f}
    &= \frac{\partial}{\partial r_f} \left(1-\sum_{h\in S_s}r_h^2 -2\sum_{c\in S_+}r_c^2\right)
    = -2r_f\;,\\ \nonumber\\
    \frac{\partial P_0}{\partial r_a} &= \frac{\partial}{\partial r_a} \left(1-\sum_{h\in S_s}r_h^2 -2\sum_{c\in S_+}r_c^2\right)
    = -4r_a\;,\\ \nonumber\\
    \frac{\partial P_0}{\partial \phi_a} &= \frac{\partial}{\partial \phi_a} \left(1-\sum_{h\in S_s}r_h^2 -2\sum_{c\in S_+}r_c^2\right)
    = 0\;.\\ \nonumber\\
    \frac{\partial P_h}{\partial r_f} &= \frac{\partial}{\partial r_f} r_h^2
    = 2r_h\delta_{h,f}\;.\\ \nonumber\\
    \frac{\partial P_h}{\partial r_a}&=0\;.
\end{align}
\begin{align}
    \frac{\partial P_h}{\partial \phi_a}&=0\;.\\ \nonumber\\
    \frac{\partial P_c}{\partial r_f} &= \frac{\partial}{\partial r_f}r_c^2\left(1+\cos(\Delta_c)\right) = 0\;.\\ \nonumber\\
    \frac{\partial P_c}{\partial r_a} &= \frac{\partial}{\partial r_a}r_c^2\left(1+\cos(\Delta_c)\right) = 2r_c\delta_{a,c}\left(1+\cos(\Delta_c)\right)\;.
\end{align}
Considering Eq. \eqref{Delta_a_main},
\begin{align}
    \frac{\partial P_c}{\partial \phi_a} &= \frac{\partial}{\partial \phi_a}r_c^2\left(1+\cos(\Delta_c)\right) = -2r_c^2\sin(\Delta_c)\delta_{a,c}\;.\\ \nonumber\\
    \frac{\partial P_{\ominus c}}{\partial r_f} &= \frac{\partial}{\partial r_f}r_c^2\left(1-\cos(\Delta_c)\right) = 0\;.\\ \nonumber \\
    \frac{\partial P_{\ominus c}}{\partial r_a} &= \frac{\partial}{\partial r_a}r_c^2\left(1-\cos(\Delta_c)\right) = 2r_c\delta_{a,c}\left(1-\cos(\Delta_c)\right)\;.\\ \nonumber \\
    \frac{\partial P_{\ominus c}}{\partial \phi_a} &= \frac{\partial}{\partial \phi_a}r_c^2\left(1-\cos(\Delta_c)\right) = 2r_c^2\sin(\Delta_c)\delta_{a,c}\;.
\end{align}    

Putting all these derivatives in Eq. \eqref{I_expanded}, we can calculate $\mathcal{I}$ by blocks. We have:
\begin{align}
    \mathbf{A}_{f,g}:=&\; \mathcal{I}_{r_fr_g}\nonumber\\
    =&\; 4\frac{r_fr_g}{r_0^2}+\sum_{h\in S_s}\frac{1}{r_h^2}4r_gr_f\delta_{h,f}\delta_{h,g}+0\nonumber\\
    =&\; 4\frac{r_fr_g}{r_0^2}+\frac{1}{r_f^2}4r_gr_f\delta_{f,g}\nonumber\\
    =&\; 4\frac{r_fr_g}{r_0^2}+4\delta_{f,g}\;.
\end{align}
Also,
\begin{equation}
    \mathbf{B}_{f,a}:=\mathcal{I}_{r_fr_a} = 8\frac{r_fr_a}{r_0^2} + 0 + 0 = 8\frac{r_fr_a}{r_0^2}\;,
\end{equation}
and
\begin{equation}
    \mathcal{I}_{r_f\phi_a} = 0 + 0 + 0 = 0\;,
\end{equation}
similarly 
\begin{align}
    \mathbf{C}_{a,b}:=&\; \mathcal{I}_{r_ar_b}\nonumber\\
    =&\; 16\frac{r_ar_b}{r_0^2} + 0 \nonumber\\
    & +\sum_{c\in S_+}\left(\frac{1}{r_c^2(1+\cos\Delta_c)}4r_c^2(1+\cos\Delta_c)^2\delta_{a,c}\delta_{b,c} + \frac{1}{r_c^2(1-\cos\Delta_c)}4r_c^2(1-\cos\Delta_c)^2\delta_{a,c}\delta_{b,c}\right)\nonumber\\
    =&\; 16\frac{r_ar_b}{r_0^2} + \left( 4(1+\cos\Delta_a) + 4(1-\cos\Delta_a) \right)\delta_{a,b}\nonumber\\
    =&\; 16\frac{r_ar_b}{r_0^2} + 8\delta_{a,b}\;,
\end{align}
and
\begin{align}
    \mathcal{I}_{r_a\phi_b} =&\; 0 + 0 \nonumber\\
    &+ \sum_{c\in S_+}\left( \frac{-4}{r_c^2(1+\cos\Delta)}r_c^3(1+\cos\Delta_c)\sin\Delta_c\delta_{a,c}\delta_{b,c} + \frac{4}{r_c^2(1-\cos\Delta_c)}r_c^3(1-\cos\Delta_c)\sin\Delta_c\delta_{a,c}\delta_{b,c}\right)\nonumber\\
    =&\; \sum_{c\in S_+}\left(-4r_c\sin\Delta_c + 4r_c\sin\Delta_c\right)\delta_{a,c}\delta_{b,c}\nonumber\\
    =&\; 0\;,
\end{align}
finally
\begin{align}
    \mathbf{D}_{a,b}:=&\; \mathcal{I}_{\phi_a\phi_b}\nonumber\\
    =&\; 0 + 0 +\sum_{c\in S_+}\left( \frac{4}{r_c^2(1+\cos\Delta_c)}r_c^4\sin^2\Delta_c\delta_{a,c}\delta_{b,c} + \frac{4}{r_c^2(1-\cos\Delta_c)}r_c^4\sin^2\Delta_c\delta_{a,c}\delta_{b,c}\right) \nonumber\\
    =&\; \sum_{c\in S_+} 4r_c^2\sin^2\Delta_c \delta_{a,c}\delta_{b,c} \left( \frac{1}{1+\cos\Delta_c}+ \frac{1}{1-\cos\Delta_c}\right)\nonumber\\
    =&\; \sum_{c\in S_+} 4r_c^2\sin^2\Delta_c \delta_{a,c}\delta_{b,c} \left( \frac{1-\cos\Delta_c}{1-\cos^2\Delta_c}+ \frac{1+\cos\Delta_c}{1-\cos^2\Delta_c}\right)\nonumber\\
    =&\; \sum_{c\in S_+} 8r_c^2\delta_{a,c}\delta_{b,c}\nonumber\\
    =&\; 8r_a^2\delta_{a,b}\;.
\end{align}

Summarizing, we obtained the following classical Fisher information matrix for our procedure:
\begin{equation}
    \mathcal{I} = \begin{pmatrix}
        \mathbf{A} & \mathbf{B} & \mathbf{0}\\
        \mathbf{B}^T & \mathbf{C} & \mathbf{0}\\
        \mathbf{0} & \mathbf{0} & \mathbf{D}
    \end{pmatrix}\;,
\end{equation}
where each block has the same definition of the corresponding block in the quantum Fisher information matrix in Eq. \eqref{F_even}.

\section{Comparison of classical and quantum Fisher information matrices slightly away from the identity}
\label{AppX}

\subsection{Alternative Parametrization of $U$: Gell-Mann matrices}

In order to compare the classical and quantum Fisher information matrices, we firstly need to guarantee the unitarity of $U$ while mantaining the independence of the parameters. A suitable way to do this is by writing an arbitrary unitary gate as the exponential of a Hermitian operator $H$:
\begin{equation}
\label{U_exp}
    U=\exp(iH)\;.
\end{equation}
We can always write $H$ as
\begin{equation}
\label{Hamiltonian}
    H=\sum_{j=1}^{d^2-1}\lambda_jT_j\;,
\end{equation}
where $T_j$ are the generalized Gell-Mann matrices and $\lambda_j$ are $d^2-1$ real parameters (see definition in Ref. \cite{Bertlmann_2008}).

The set $\{T_j\}$ of the $d^2-1$ generalized Gell-Mann matrices plus the operator $\sqrt{2/d}\; I$ is a set of $d^2$ linear operators satisfying the orthogonality relation $Tr[T_i^\dagger T_j]=2\delta_{i,j}$, with $\delta_{i,j}$ being the Kronecker delta. Hence, this set is an orthogonal basis for the space of $d$-dimensional linear operators. As a consequence, any operator $U$ can be spanned as:
\begin{equation}
\label{U_spaned_Gellmann}
    U=\sum_{k=1}^{d^2}u_k^{(GM)}T_k\;,
\end{equation}
where $u_k^{(GM)}$ are complex coefficients. In order to relate this expansion with the one in the main text, we will rewrite here Eq.~\eqref{Weyl-Heisenberg expansion} as:
\begin{equation}
    U=\sum_{k=1}^{d^2}u_k^{(WH)}D_k\;,
\end{equation}
where $u_k^{(WH)}$ are complex coefficients and $D_k$ are the Weyl-Heisenberg operators.

Notice that every Gell-Mann matrix can also be spaned in the Weyl-Heisenberg basis. We can write
\begin{equation}
    T_k=\sum_{j=1}^{d^2}t^{(k)}_j D_j\;,
\end{equation}
where the $j$-th component $t^{(k)}_j$ of the $k$-th Gell-Mann matrix can be calculated as\begin{equation}
    t^{(k)}_i=\frac{1}{d}Tr[D_i^\dagger T_k]\;.
\end{equation}

Let us highlight that now we have two orthogonal bases in the space of $d$-dimensional operators. Moreover, we know from Eq.~\eqref{U-normalization} that every unitary transformation is associated to a unitary vector when it is expanded in the Weyl-Heisenberg basis. This is not the case if we use the Gell-Mann basis, as can be easily seen by taking the identity, which in the Gell-Mann basis has a unique component different from zero with value $\sqrt{d/2}$. We would like to normalize Gell-Mann matrices in such a way that any unitary operator is still associated to a unitary vector. To this end, we define the normalized Gell-Mann matrices as
\begin{equation}
    \tilde{T}_k=\sqrt{\frac{d}{2}}T_k\;,
\end{equation}
whose components in the Weyl-Heisenberg basis are
\begin{equation}
    \tilde{t}^{(k)}_i=\frac{1}{d}Tr[D_i^\dagger \tilde{T}_k]=\sqrt{\frac{d}{2}}t^{(k)}_i\;.
\end{equation}

These components define the vectors of an alternative orthonormal basis in a $d^2$-dimensional vector space. Indeed, let us calculate the inner product of the $k$-th vector by the $j$-th vector:
\begin{align}
    \sum_{i}\tilde{t}^{(k)}_i\cdot \tilde{t}^{(j)*}_i&=\sum_i \frac{1}{d^2}\cdot Tr[D_i^\dagger \tilde{T}_k]\cdot Tr[D_i^\dagger \tilde{T}_j]^*\nonumber\\
    &=\sum_i\frac{1}{d^2}\cdot\frac{d}{2} \cdot Tr[D_i^\dagger T_k]\cdot Tr[D_i^\dagger T_j]^*\nonumber\\
    &=\frac{1}{2d}\sum_{i,m,n}\langle m|D_i^\dagger T_k|m\rangle\langle n|D_i^\dagger T_j|n\rangle^*\nonumber\\
    &=\frac{1}{2d}\sum_{i,m,n}\langle m|D_i^\dagger T_k|m\rangle\langle n|T_j^\dagger D_i|n\rangle\nonumber\\
    &=\frac{1}{2d}\sum_{i,m,n}\langle n|T_j^\dagger D_i|n\rangle\langle m|D_i^\dagger T_k|m\rangle\nonumber\\
    &=\frac{1}{2d}\sum_{m,n}\langle n|T_j^\dagger \sum_i\left(D_i|n\rangle\langle m|D_i^\dagger\right) T_k|m\rangle\;.
\end{align}
But notice that
\begin{align}
    \sum_i\left(D_i|n\rangle\langle m|D_i^\dagger\right)&=\sum_{i_x,i_z}\left(X^{i_x}Z^{i_z}|n\rangle\langle m|Z^{-i_z}X^{-i_x}\right)\nonumber\\
    &=\sum_{i_x,i_z}\omega^{i_z(n-m)}|n\oplus i_x\rangle\langle m\oplus i_x|\nonumber\\
    &=\sum_{i_x}d\cdot \delta_{m,n}|n\oplus i_x\rangle\langle m\oplus i_x|\;.
\end{align}
Then,
\begin{align}
    \sum_{i}\tilde{t}^{(k)}_i\cdot \tilde{t}^{(j)*}_i&=\frac{1}{2d}\sum_{m,n,i_x}d\cdot \delta_{m,n}\langle n|T_j^\dagger |n\oplus i_x\rangle\langle m\oplus i_x| T_k|m\rangle\nonumber\\
    &=\frac{1}{2}\sum_{m}\langle m|T_j^\dagger \sum_{i_x}\left(|m\oplus i_x\rangle\langle m\oplus i_x|\right) T_k|m\rangle\nonumber\\
    &=\frac{1}{2}\sum_{m}\langle m|T_j^\dagger T_k|m\rangle\nonumber\\
    &=\frac{1}{2}\cdot Tr[T_j^\dagger T_k]\nonumber\\
    &=\frac{1}{2}\cdot 2\delta_{j,k}\nonumber\\
    &=\delta_{j,k}\;.
\end{align}

In the next subsection, we will use $\{\tilde{t}^{(k)}\}$ as the measurement basis for the control system in our circuit.

\subsection{Estimating the parameteres of $U$ (first order)}

Let us consider an arbitrary unitary transformation $U$ close to the identity. If we expand Eq.~\eqref{U_exp} to first order, we get
\begin{equation}
    U\approx I+iH\;.
\end{equation}
Using Eq.~\eqref{Hamiltonian},
\begin{equation}
    U\approx I+i\sum_{j=1}^{d^2-1}\lambda_jT_j\;.
\end{equation}
Considering the normalized Gell-Mann matrices, we have 
\begin{equation}
    U\approx I+i\sqrt{\frac{2}{d}}\sum_{j=1}^{d^2-1}\lambda_j\tilde{T}_j\;,
\end{equation}
and we can rewrite Eq.~\eqref{U_spaned_Gellmann} as
\begin{align}
    U&=\sqrt{\frac{2}{d}}\sum_{k=1}^{d^2}u_k^{(GM)}\tilde{T}_k\nonumber\\
    &=\sum_{k=1}^{d^2}\tilde{u}_k^{(GM)}\tilde{T}_k\;,
\end{align}
with $\tilde{u}_k^{(GM)}=\sqrt{2/d}\cdot u_k^{(GM)}$ being the components of the unitary vector representing $U$ in the normalized Gell-Mann basis. Thus, we have:
\begin{align}
    \tilde{u}_0^{(GM)}&\approx 1\;,\\
    \tilde{u}_j^{(GM)}&\approx i\sqrt{\frac{2}{d}}\lambda_j\;\text{, for }1\leq j \leq d^2-1\;.
\end{align}

Now let us consider again our circuit. By measuring the final state of the control system in the basis induced by the normalized Gell-Mann matrices, we get outcome probabilities
\begin{align}
    p_0&=|\tilde{u}_0^{(GM)}|^2\approx 1\;,\\
    p_j&=|\tilde{u}_j^{(GM)}|^2\approx \frac{2\lambda_j^2}{d}\;\text{, for }1\leq j \leq d^2-1\;,
\end{align}
from where we get an estimation of the parameters. Assuming $\lambda_j\geq 0$, we have
\begin{equation}
\label{estimator}
    \lambda_j\approx \sqrt{\frac{d\cdot p_j}{2}}\;.
\end{equation}
Now we can use this estimator to assess the quality of the approximation when we go slightly away from the identity, and also compare numerically the classical and quantum Fisher information matrices.

\subsection{Quality of the estimation}
\begin{figure*}[t]
    \centering
    \begin{tabular}{cc}
    \includegraphics[width=80mm]{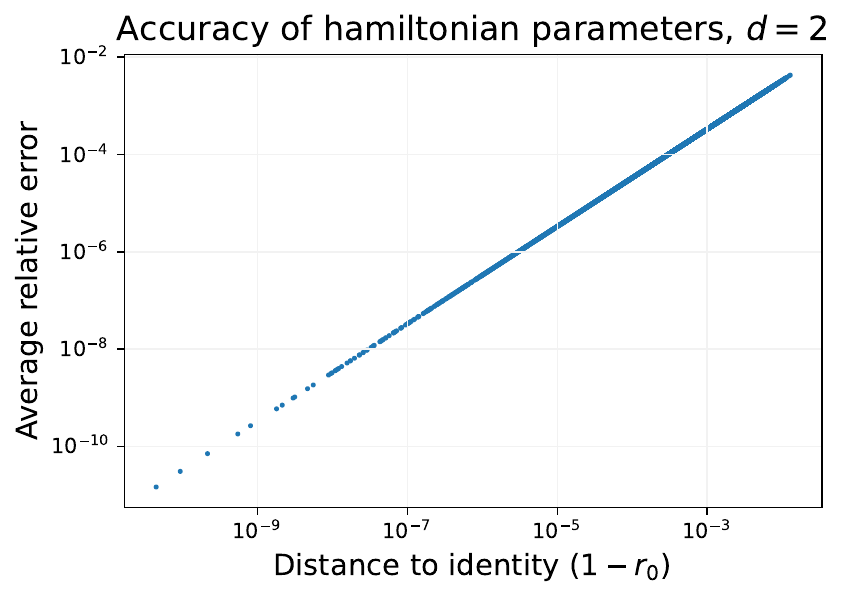} &
     \includegraphics[width=80mm]{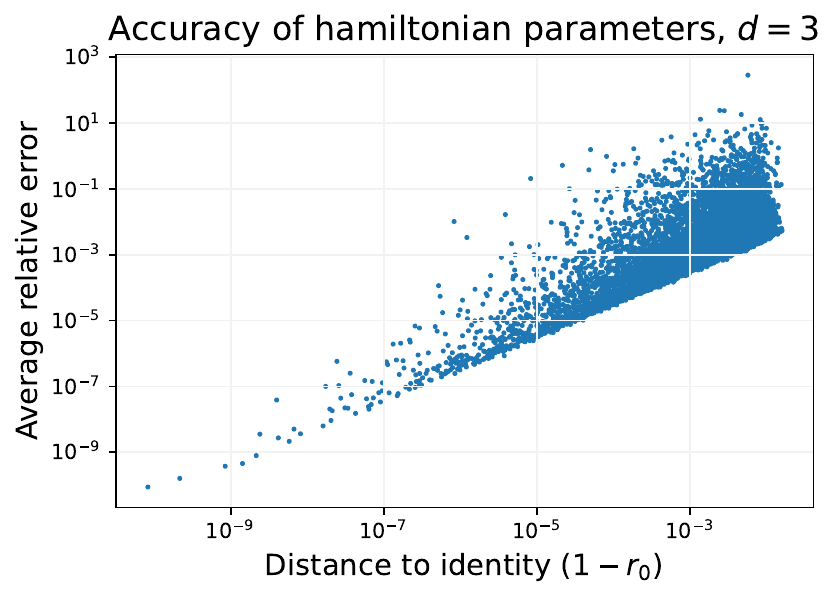}  \\
     (a) & (b) \\
     \includegraphics[width=80mm]{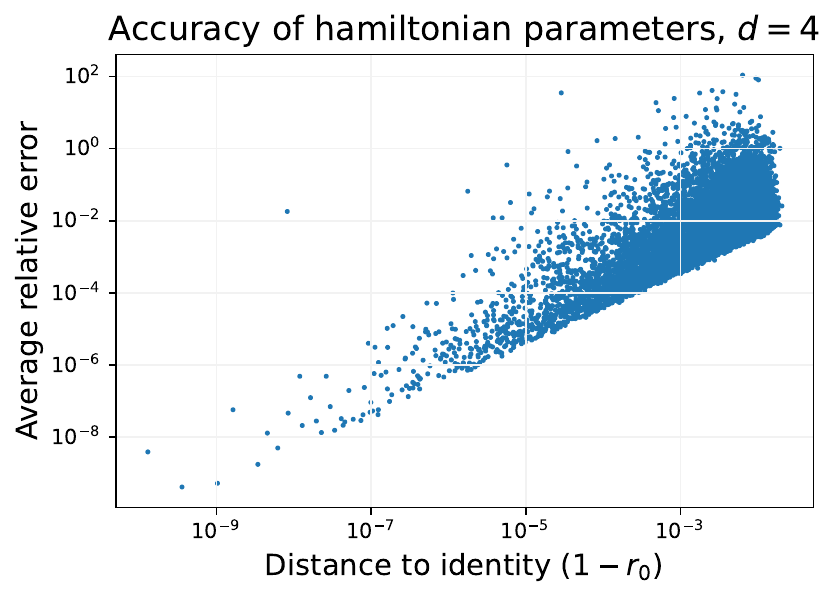} &
     \includegraphics[width=80mm]{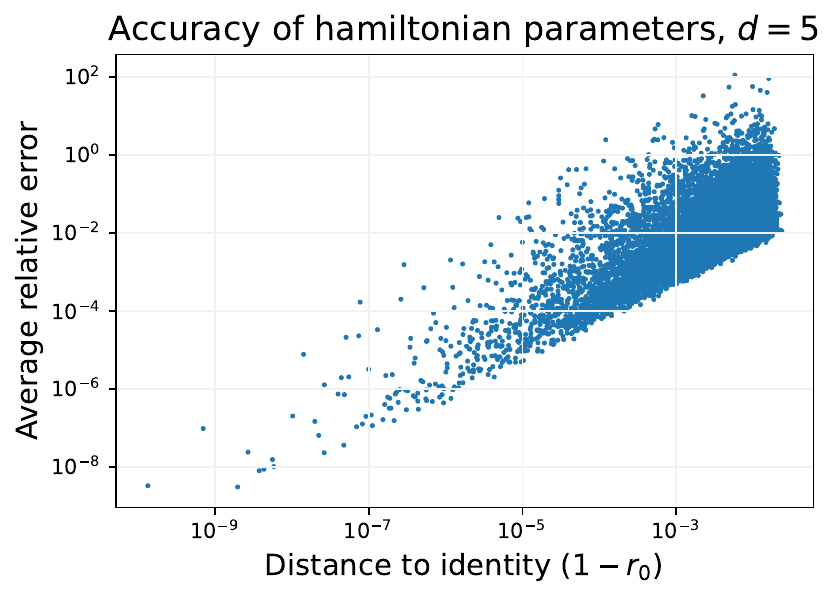} \\
     (c) & (d)
    \end{tabular}
    \caption{Average relative error of estimated Hamiltonian parameters using first order approximation. Figures (a), (b), (c) and (d) show unitary transformations close to the identity in dimensions 2, 3, 4 and 5, respectively.}
    \label{Accuracy plots}
    \end{figure*}

We assessed numerically the accuracy of the parameter estimation using Eq.~\eqref{estimator}. For each dimension $d=2,3,4$ and $5$ we created 10000 random unitary transformations close to the identity. Hamiltonian parameters were chosen randomly within a variable range up to $[0, 0.1]$, in such a way that we could scan different distances to the identity. Although it is not properly a distance, we used $1-r_0$ as a measure of how close is $U$ to the identity operator, with $r_0$ being the component of $U$ in the Weyl-Heisenberg basis as in the main text (we also tried with the trace distance, obtaining equivalent results). For each unitary transformation, we estimated the parameters in the asymptotic limit using Eq.~\eqref{estimator}. The average relative error of the estimated parameters for each unitary is shown in Fig. \ref{Accuracy plots}. As expected, the accuracy of the estimation gets worse as $U$ is further from the identity. For dimensions 3, 4 and 5 the approximation leads to an average relative error easily surpassing $1\%$ when $1-r_0>1\times 10^{-3}$, i.e. for $r_0<0.999$. For dimension $d=2$, the average relative error remains below $1\%$ in the range of parameters considered.

\subsection{Classical and quantum Fisher Information comparison}

\begin{figure*}[t]
    \centering
    \begin{tabular}{cc}
    \includegraphics[width=80mm]{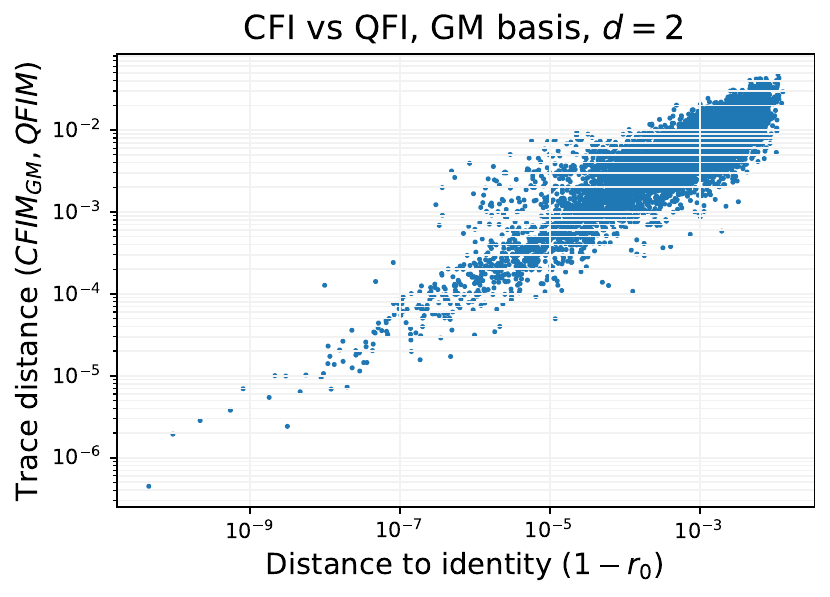} &
     \includegraphics[width=80mm]{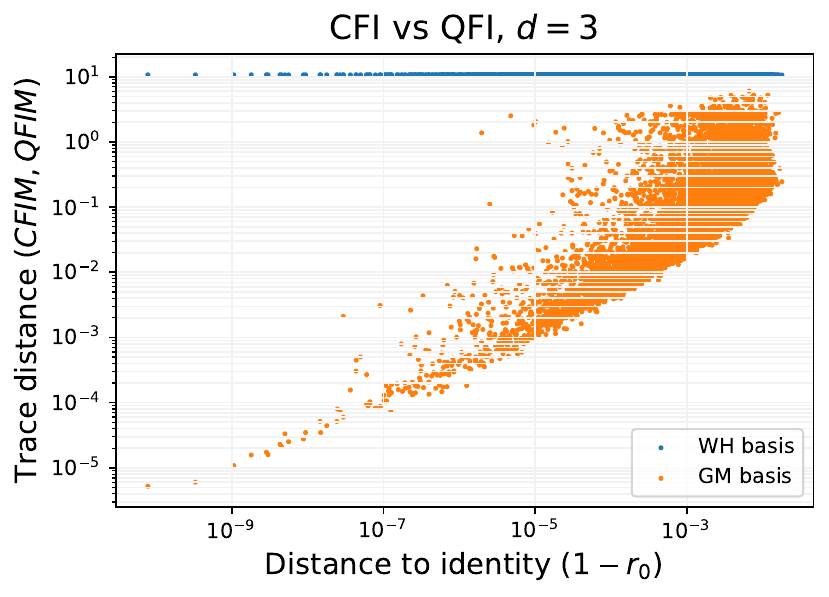}  \\
     \includegraphics[width=80mm]{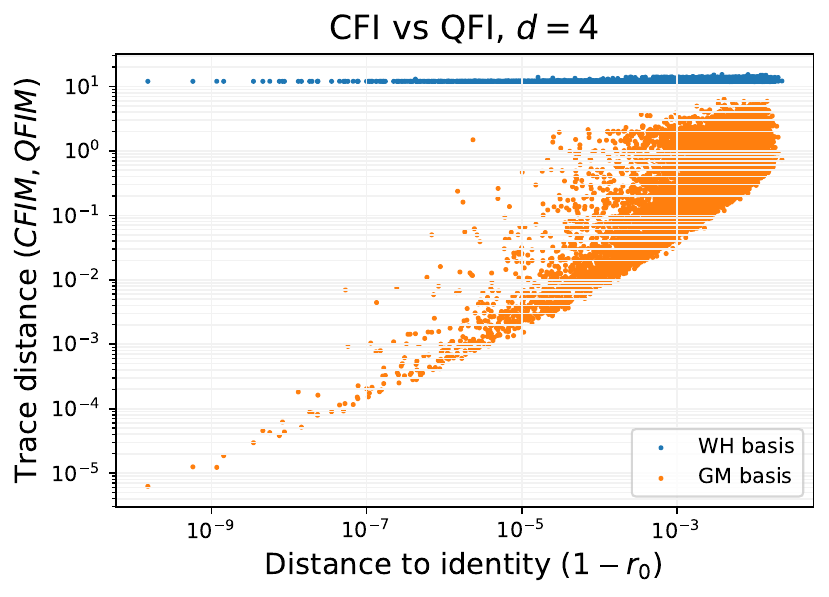} &
     \includegraphics[width=80mm]{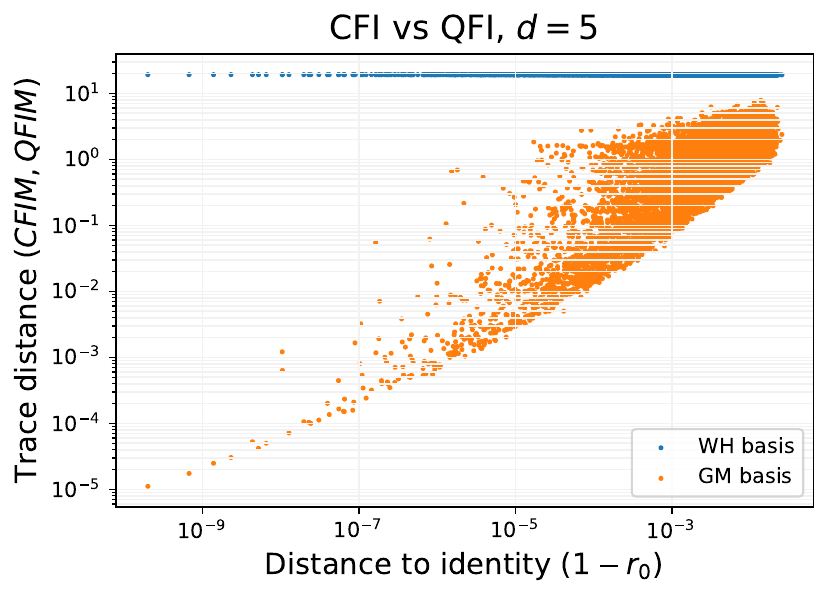} \\
    \end{tabular}
    \caption{Trace distance between classical Fisher information (CFI) and quantum Fisher information (QFI) matrices. For dimension 3, 4 and 5, classical Fisher information is computed from probabilities obtained by measuring the control system in two different bases: the computational basis (WH, blue dots) and the basis induced by normalized Gell-Mann matrices (GM, orange dots). For dimension $d=2$ both bases are exactly the same.}
    \label{CFI_plots}
\end{figure*}

For the same unitaries as the previous section, we calculated the quantum Fisher information matrix respect to the Hamiltonian parameters and the classical Fisher information matrix for two different measurement schemes: 1) when the control system of our circuit is measured in the computational basis (corresponding to the Weyl-Heisenberg basis in the space of linear operators) and 2) in the basis induced by the normalized Gell-Mann matrices. We used the trace distance between the quantum an classical Fisher information matrices as figure of merit. Fig \ref{CFI_plots} shows that for dimensions 3, 4 and 5 the classical Fisher information matrix is closer to the quantum Fisher information matrix when the control system is measured in the basis induced by Gell-Mann matrices. Moreover, this distance approaches to zero as $U$ approaches the identity, while the distance between the classical Fisher information using the computational basis and quantum Fisher information matrices remains constant when $U$ approaches to identity. In the case of dimension $d=2$, where both bases are the same, the trace distance achieves lower values than in higher dimensions.

\subsection{Far from the identity}

The estimation of parameters performed above only considers a first order approximation of $U$ around the identity. It would be interesting to go beyond this approximation to estimate parameters far from the identity. For example, we could approximate $U$ to second order and calculate the Hamiltonian parameters from the outocome probabilities by using equations similar to Eq.~\eqref{estimator}. However, we could not find neither analytical nor numerical solutions for the nonlinear equations system not even in dimension $d=3$.
An alternative approach consists in having a good guess for the unitary gate (let us call it $\tilde{U}$) and expand $U$ to first order around $\tilde{U}$. However, this approach is not straightforward, since it involves derivatives of operators which do not necessarily commute. Finally, it would be interesting to explore different parametrizations for unitary operators, what we let for future work.

\section{Qubit case without prior information}
\label{no_prior_information}

Any qubit unitary gate can be written as \eqref{u_main}, which is expanded as
\begin{align}
    U&=\cos(\alpha)I-i\sin(\alpha)\hat{n}\cdot\hat{\sigma}\nonumber\\
    &=c_II-ic_xX-ic_yY-ic_zZ\nonumber\\
\label{Uqubits}
    &=c_II-ic_xX+c_yXZ-ic_zZ\;,
\end{align}
where $c_I=\cos(\alpha)$ and $c_k=\sin(\alpha)n_k$ for $k\in\{x,y,z\}$, with $\alpha \in [0,\pi/2]$ and $\hat{n} \in \mathbb{R}^3$ a unit vector.

\begin{figure}[h]
    \centering
    \begin{quantikz}
    \lstick[wires=1]{$|\psi\rangle_0$} \slice{$|\Phi^0\rangle_{012}$} & \qw & \gate[wires=1]{X} & \gate[wires=1]{Z} & \gate[wires=1]{U} & \gate[wires=1]{Z} & \gate[wires=1]{X} & \qw \slice{$|\Phi^7\rangle_{012}$} & \gate[wires=1]{X} & \gate[wires=1]{Z} & \qw \rstick[wires=1] {$|\psi\rangle_0$}\\
    \lstick[wires=1]{$|0\rangle_1$} & \gate[wires=1]{H} & \qw & \octrl{-1} & \qw & \ctrl{-1} & \qw & \gate[wires=1]{H} & \octrl{-1} & \qw & \qw \rstick[wires=2]{$|\varphi^Z\rangle_{12}$}\\
    \lstick[wires=1]{$|0\rangle_2$} & \gate[wires=1]{H} & \ctrl{-2} & \qw & \qw & \qw & \octrl{-2} & \gate[wires=1]{H} & \qw & \ctrl{-2} & \qw
    \end{quantikz}
    \caption{Measurement in $Z$ eigenbasis. The circuit illustrates the qubit version of the circuit in Fig. \ref{Circuit0}, without the last $X$ gate in the second control qubit.}
    \label{circuit_z}
\end{figure}
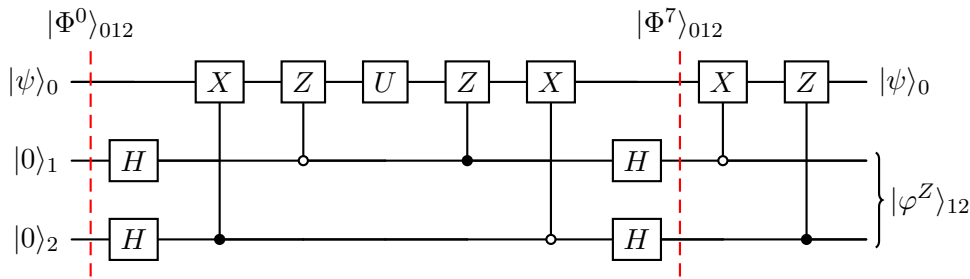

Notice that for the case of qubits the quantum state $|\Phi^9\rangle_{012}$ in Eq.~\eqref{phi9} becomes 
\begin{align}
    |\Phi^9\rangle_{012}&= |\psi\rangle_0\otimes\bigg(-ic_z|00\rangle_{12}+c_I|01\rangle_{12}+c_y|10\rangle_{12}-ic_x|11\rangle_{12}\bigg)\nonumber\\
    &=|\psi\rangle_0\otimes|\varphi^Z\rangle_{12}\;,
\end{align}
where $|\varphi^Z\rangle_{12}=-ic_z|00\rangle_{12}+c_I|01\rangle_{12}+c_y|10\rangle_{12}-ic_x|11\rangle_{12}\;$. Measuring this state in the computational basis (see Fig. \ref{circuit_z}) leads to probabilities  $P^Z_{00} = |c_z|^2$, $P^Z_{01} = |c_I|^2$, $P^Z_{10} = |c_y|^2$ and $P^Z_{11} = |c_x|^2$.

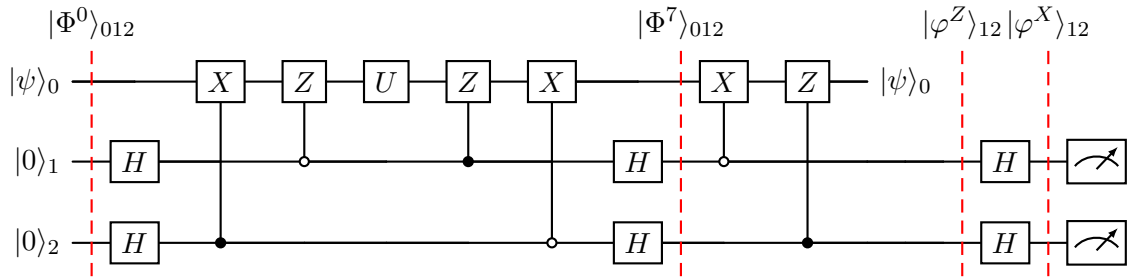
\begin{figure}[h]
    \centering
    \begin{quantikz}
    \lstick[wires=1]{$|\psi\rangle_0$} \slice{$|\Phi^0\rangle_{012}$} & \qw & \gate[wires=1]{X} & \gate[wires=1]{Z} & \gate[wires=1]{U} & \gate[wires=1]{Z} & \gate[wires=1]{X} & \qw \slice{$|\Phi^7\rangle_{012}$} & \gate[wires=1]{X} & \gate[wires=1]{Z} & \qw \rstick[wires=1] {$|\psi\rangle_0$} & \setwiretype{n}& \slice{$|\varphi^Z\rangle_{12}$} & \slice{$|\varphi^X\rangle_{12}$} &\\
    \lstick[wires=1]{$|0\rangle_1$} & \gate[wires=1]{H} & \qw & \octrl{-1} & \qw & \ctrl{-1} & \qw & \gate[wires=1]{H} & \octrl{-1} & \qw & \qw & \qw & \qw & \gate[wires=1]{H} & \meter{}\\
    \lstick[wires=1]{$|0\rangle_2$} & \gate[wires=1]{H} & \ctrl{-2} & \qw & \qw & \qw & \octrl{-2} & \gate[wires=1]{H} & \qw & \ctrl{-2} & \qw & \qw & \qw & \gate[wires=1]{H} & \meter{}
    \end{quantikz}
    \caption{Measurement in $X$ eigenbasis. The circuit illustrates the qubit version of the circuit in Fig. \ref{Circuit0}, without the last $X$ gate in the second control qubit. Additional rotation is performed in each control qubit in order to measure them in the $X$ eigenbasis.}
    \label{circuit_x}
\end{figure}

\begin{figure}[h]
    \centering
    \begin{quantikz}
    \lstick[wires=1]{$|\psi\rangle_0$} \slice{$|\Phi^0\rangle_{012}$} & \qw & \gate[wires=1]{X} & \gate[wires=1]{Z} & \gate[wires=1]{U} & \gate[wires=1]{Z} & \gate[wires=1]{X} & \qw \slice{$|\Phi^7\rangle_{012}$} & \gate[wires=1]{X} & \gate[wires=1]{Z} & \qw \rstick[wires=1] {$|\psi\rangle_0$} & \setwiretype{n} & \slice{$|\varphi^Z\rangle_{12}$} & &  \slice{$|\varphi^Y\rangle_{12}$} &\\
    \lstick[wires=1]{$|0\rangle_1$} & \gate[wires=1]{H} & \qw & \octrl{-1} & \qw & \ctrl{-1} & \qw & \gate[wires=1]{H} & \octrl{-1} & \qw & \qw & \qw & \qw & \gate[wires=1]{S} & \gate[wires=1]{H} & \meter{}\\
    \lstick[wires=1]{$|0\rangle_2$} & \gate[wires=1]{H} & \ctrl{-2} & \qw & \qw & \qw & \octrl{-2} & \gate[wires=1]{H} & \qw & \ctrl{-2} & \qw & \qw & \qw & \gate[wires=1]{S} & \gate[wires=1]{H} & \meter{}
    \end{quantikz}
    \caption{Measurement in $Y$ eigenbasis. The circuit illustrates the qubit version of the circuit in Fig. \ref{Circuit0}, without the last $X$ gate in the second control qubit. Additional rotation is performed in each control qubit in order to measure them in the $Y$ eigenbasis.}
    \label{circuit_y}
\end{figure}
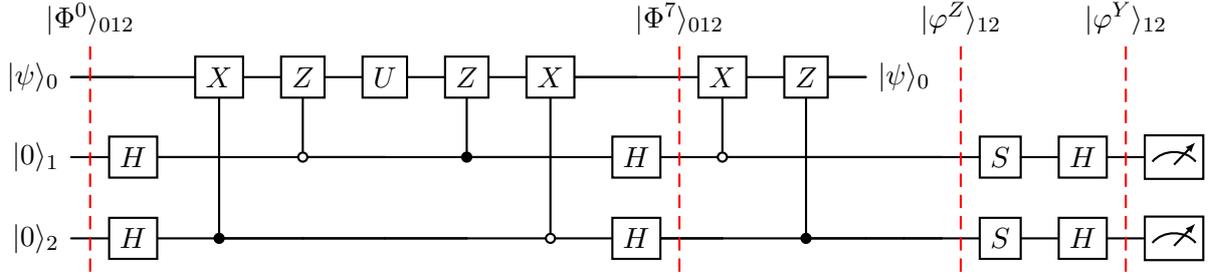

For measuring the control qubits in the $X$ basis we need to apply a Hadamard gate on each one  (see Fig. \ref{circuit_x}). We get the new state
\begin{align}
    |\varphi^X\rangle_{12}=&\; \frac{1}{2}\bigg[ (-ic_z+c_I+c_y-ic_x)|00\rangle_{12}+(-ic_z-c_I+c_y+ic_x)|01\rangle_{12}\nonumber\\
    & +(-ic_z+c_I-c_y+ic_x)|10\rangle_{12}+(-ic_z-c_I-c_y-ic_x)|11\rangle_{12}\bigg]\;,
\end{align}
leading to probabilities
\begin{align}
    P_{00}^{X}&=\frac{1}{4}\bigg( (c_I+c_y)^2+(c_x+c_z)^2\bigg), \nonumber\\
    P_{01}^{X}&=\frac{1}{4}\bigg( (c_I-c_y)^2+(c_x-c_z)^2\bigg), \nonumber\\
    P_{10}^{X}&=\frac{1}{4}\bigg( (c_I-c_y)^2+(c_x-c_z)^2\bigg),\nonumber\\
    P_{11}^{X}&=\frac{1}{4}\bigg( (c_I+c_y)^2+(c_x+c_z)^2\bigg)\;.
    \label{probs1f}
\end{align}
In order to measure the qubits in the $Y$ eigenbasis, we apply a $\pi/2$-phase gate, defined as
\begin{equation}
    S=|0\rangle\langle 0|+i|1\rangle\langle 1|\;,
\end{equation}
followed by a Hadamard gate on each control qubit (see Fig. \ref{circuit_y}). We get the state:
\begin{align}
    |\varphi^Y\rangle_{12}=&\; \frac{i}{2}\bigg[ (-c_z+c_I+c_y+c_x)|00\rangle_{12}+(-c_z-c_I+c_y-c_x)|01\rangle_{12}\nonumber\\
    & +(-c_z+c_I-c_y-c_x)|10\rangle_{12}+(-c_z-c_I-c_y+c_x)|11\rangle_{12}\bigg]\;,
\end{align}
and the outcome probabilities
\begin{align}
    P_{00}^{Y}&=\frac{1}{4}\bigg( c_I+c_y+c_x-c_z\bigg)^2, \nonumber\\
    P_{01}^{Y}&=\frac{1}{4}\bigg(-c_I+c_y-c_x-c_z\bigg)^2, \nonumber\\
    P_{10}^{Y}&=\frac{1}{4}\bigg( c_I-c_y-c_x-c_z\bigg)^2, \nonumber\\
    P_{11}^{Y}&=\frac{1}{4}\bigg( -c_I-c_y+c_x-c_z\bigg)^2\;.
    \label{probs2f}
\end{align}

Since the absolute values of $c_I$, $c_x$, $c_y$ and $c_z$ are known from the first experiment, we can now use the outcome probabilities \eqref{probs1f} and \eqref{probs2f} to discriminate their signs. We denote $r_i\equiv|c_i|$, $s_i\equiv \text{sgn}(c_i)$ and $s_{xz}=s_x\cdot s_z$. We obtain
\begin{equation}
    P_{00}^{X}+P_{11}^{X}-P_{00}^{Y}-P_{11}^{Y}=2c_xc_z\;.
\end{equation}
Hence
\begin{equation}
\label{sxz}
    s_{xz}=\text{sgn}(P_{00}^{X}+P_{11}^{X}-P_{00}^{Y}-P_{11}^{Y})\;.
\end{equation}
Also, it is easy to show that
\begin{equation}
    P_{00}^{X}+P_{11}^{X}+P_{00}^{Y}+P_{11}^{Y}=1+2c_Ic_y\;,
\end{equation}
from where we get
\begin{equation}
\label{sy}
    s_y=\text{sgn}(P_{00}^{X}+P_{11}^{X}+P_{00}^{Y}+P_{11}^{Y}-1)\;,
\end{equation}
where we have used the fact that $c_I^2+c_x^2+c_y^2+c_z^2=1$ and $c_I>0$.

Now we have two ways to obtain $s_x$. Firstly, noticing that
\begin{equation}
    P_{10}^{Y}+P_{11}^{Y}=\frac{1}{2}-c_xc_I+c_zc_y
\end{equation}
and that $s_z=s_xs_{xz}$, we can write
\begin{equation}
    P_{10}^{Y}+P_{11}^{Y}=\frac{1}{2}-s_xr_xr_I+s_xs_{xz}s_yr_zr_y\;,
\end{equation}
from where we get
\begin{equation}
\label{sx1}
    s_x=\frac{P_{10}^{Y}+P_{11}^{Y}-1/2}{s_{xz}s_yr_zr_y-r_xr_I}\;.
\end{equation}
Another possibility is start from
\begin{equation}
    P_{10}^{Y}-P_{11}^{Y}=c_yc_x-c_zc_I+c_xc_z-c_Ic_y
\end{equation}
to get
\begin{equation}
\label{sx2}
    s_x=\frac{P_{10}^{Y}-P_{11}^{Y}+s_yr_Ir_y-s_{xz}r_xr_z}{s_yr_yr_x-s_{xz}r_Ir_z}\;.
\end{equation}

Finally, we obtain $s_z=s_{xz}\cdot s_x$.

Notice that the Eqs. \eqref{sx1} and \eqref{sx2} are ill defined in some cases. Then, the complete protocol to estimate the unitaries from the outcome probabilities must consider special cases. We summarize the discrimination procedure in the following steps:

\begin{enumerate}
    \item Get the magnitude of the coefficients $r_i$ from $P^Z_{jk}$.
    \item If three of the $r_i$ are zero, then $U\in\{I,X,Y,Z\}$.
    \item If $r_z=r_x=0$, we only need to find $s_y$ using Eq. \eqref{sy}. In case it is exactly zero, it means that either $r_I$ or $r_y$ is close to zero; we choose $s_y$ randomly.
    \item If $r_I=r_y=0$, we get freedom in the global phase. Hence, we choose $s_x$ randomly and calculate $s_z$ from $s_{xz}$ and Eq. \eqref{sxz}.
    \item Otherwise, we calculate $s_{xz}$ and $s_x$ using Eqs. \eqref{sxz} and \eqref{sy}, respectively. Then we choose Eq. \eqref{sx1} or \eqref{sx2} according to which expression has the denominator with larger absolute value, in order to diminish the statistical variation on the estimation of $s_x$. If both denominators are zero, we choose $s_x$ as the numerator of Eq. \eqref{sx2}.
\end{enumerate}

The codes for this algorithm are shown in github \cite{github}.

\section{Noise model}
\label{app:noise}
We use the \texttt{NoiseModel} class from Qiskit to incorporate noise in the simulation of our procedure. The simulation with full noise was accomplished by importing an approximate noise model of the \texttt{ibmq\_quito} quantum processor of IBM. We consider only qubits 0, 1 and 2 from the five qubits of this processor, assigning qubit 1 for the target qubit in our protocol and qubits 0 and 2 as control register. Table \ref{tab full noise} shows the relevant error parameters for this setting, which are relaxation times T1 and T2, single-qubit gates error SX, readout or measurement errors and control-not gates error.

A second noisy simulation was performed considering only control-not gate erros. To this end, we fixed single-qubit gates and readout errors to zero. To avoid relaxation, we kept relaxation times but set the implementation time of each gate to zero. Finally, a full noise simulation but with ideal control-not gate was performed. In this case, only control-not error was consider equal to zero. Table \ref{tab fixed noise} shows the error parameters considered in these scenarios. These values are the maximum error parameters in \texttt{ibmq\_quito} at the moment of setting the model and we consider the same values for each qubit.

\begin{table}[h]
    \centering
    \begin{tabular}{|c|c|c|c|c|c|} \hline
        Error Parameter & Qubit 0 & Qubit 1 & Qubit 2 & Qubit 01 & Qubit 12 \\ \hline
         Relaxation time T1s [ns] & $87.49953 \times 10^3$ & $86.63249 \times 10^3$ & $83.6549\times 10^3$ & - & - \\ \hline
         Relaxation time T2s [ns] & $121.65781\times 10^3$ & $97.53323\times 10^3$ & $72.867\times 10^3$ & - & - \\ \hline
         Single-qubit gate error (SX) & 0.00045 & 0.0004 & 0.00027 & - & - \\ \hline
         Readout errors & 0.0406 & 0.0444 & 0.0841 & - & - \\ \hline
         Control-not error & - & - & - & 0.01021 & 0.00861 \\ \hline
    \end{tabular}
    \caption{Error parameters of \texttt{ibmq\_quito} for each qubit used in the full noise simulation.}
    \label{tab full noise}
\end{table}

\begin{table}[h]
    \centering
    \begin{tabular}{|c|c|c|c|c|c|} \hline
        Error Parameter & Value  \\ \hline
         Relaxation time T1s [ns] & $110\times 10^3$ \\ \hline
         Dephasing time T2s [ns] & $147\times 10^3$  \\ \hline
         Error probabilities (SX) & 0.00045  \\ \hline
         Readout errors & 0.0841 \\ \hline
         Control-not & 0.0142 \\ \hline
    \end{tabular}
    \caption{For simulations with noiseless control-not gate we set its value to zero. For simulations considering only a noisy control-not gate we set to zero the other noise sources.}
    \label{tab fixed noise}
\end{table}

\end{document}